\documentclass[
aps,                    %  American Physical Society
prd,                    %  Physical Review D
showpacs,               %  Displays PACS after abstract
nofootinbib,            %  Does not treat footnotes as references
showkeys,               %
preprintnumbers,        %
amsmath,               %
amssymb,               %
floatfix]               %  Fixes float errors
{revtex4}               %  REVTEX 4 Package used
\usepackage{graphicx,longtable}
\usepackage{color}
\usepackage{slashed}
\newcommand{\ba}{\begin{eqnarray}}
\newcommand{\ea}{\end{eqnarray}}
\newcommand{\ban}{\begin{eqnarray*}}
\newcommand{\ean}{\end{eqnarray*}}
\newcommand{\be}{\begin{equation}}
\newcommand{\ee}{\end{equation}}
\begin{document}

\title{SUSY radiative corrections on $\mu - \tau$ neutrino refraction including possible R-parity breaking interactions}

\author{J. Gava}
\email{gava@ipno.in2p3.fr}
\affiliation{Institut de Physique Nucl\'eaire,
 CNRS/IN2P3 UMR 8608 and Universit\'e Paris-Sud 11,
 B\^at. 100, \\ F-91406 Orsay cedex, France}

\author{C.-C. Jean-Louis}
\email{charles.jean-louis@th.u-psud.fr}
\affiliation{Laboratoire de
Physique Th\'eorique, CNRS UMR 8627, B\^at. 210, Universit\'e
Paris-Sud 11, 91405 Orsay Cedex, France}

\date{22th July 2009}

\begin{abstract}
In this paper we investigate the one-loop radiative corrections to
the neutrino indices of refraction from supersymmetric models. We
consider the Next-to Minimal Supersymmetric extension of the
Standard Model (NMSSM) which happens to be a better supersymmetric
candidate than the MSSM for both theoretical and experimental
reasons. We scan the relevant SUSY parameters and identify regions
in the parameter space which yield interesting values for
$V_{\mu\tau}$. If R-parity is broken there are significant
differences between MSSM and NMSSM contributions contrary to the
R-parity conserved case. Finally, for a non-zero CP-violating
phase, we show analytically that the presence of $V_{\mu\tau}$
will explicitly imply CP-violation effects on the supernova
electron (anti-) neutrino fluxes.
\end{abstract}


 \maketitle

%%%%%%%%%%%%%%%%%%%%%%%%%%%%%%%%%%%%%%%%%%%%%
%%%%%%%%%%%%%%%%%%%%%%%%%%%%%%%%%%%%%%%%%%%%%

\section{Introduction}

The charged current interaction between neutrinos and their
associated leptons in medium give an effective matter potential
which can lead to a resonant flavour conversion called the
Mikheyev-Smirnov-Wolfenstein (MSW)
effect~\cite{Wolfenstein:1977ue,Mikheev:1986wj}. Such behaviour
solves elegantly the solar neutrino deficit problem first pointed
out by pioneering experiment of R. Davis in the 60's
\cite{Davis:1964aa}. Such phenomenon is of crucial importance for
the propagation of neutrinos in the supernova environment whose
flux detection could yield precious information concerning the
dynamics of the density profile or fundamental neutrinos
properties like the hierarchy or the value of the third mixing
angle. In supernovae, depending on the mass hierarchy, the
electron neutrino may encounter one or two resonances via charged
current, while the muon or tau neutrinos will only interact via
neutral current, indistinguishably at the tree level in the
Standard Model (SM). Indeed, due to the absence of muon or tau
particle in such environment, coherent forward scattering may only
intervene via neutral current to which all flavours are sensitive.
Considering that matter interaction can also be seen as neutrino
index of refraction \cite{Langacker:1983bc,Lewis:1980ab}, Botella
et al. \cite{Botella:1986wy}, after showing that correction at
$O\left(\alpha \right)$ were negligible in the case of a neutral
medium \cite{Sehgal:1985bc}, proved that differences at one-loop
in the neutrino index of refraction could arise for muon and tau
neutrinos at order $O\left({\alpha\over
\pi\sin^2\theta_W}{m_\tau^2\over M_W^2}\right)$. Later,
supersymmetric radiative corrections, in the case of the minimal
supersymmetric model (MSSM) has been partially calculated
\cite{Roulet:1995ef} showing that it could give potentially much
larger radiative effects than in the SM. It is interesting to note
that such radiative corrections have been for a longtime
considered as negligible or without observable consequences for
the supernova environment. Such consideration was probably true
when not taking into account the neutrino-neutrino interaction
which has dramatically changed the vision we have of neutrino
propagating in an exploding star and which has gone through an
intense investigation
\cite{Pantaleone:1992eq,Sigl:1992fn,Samuel:1993uw,Qian:1994wh,Sigl:1994hc,
Pastor:2001iu,Pastor:2002we,Balantekin:2004ug,Duan:2005cp,Duan:2006an,Hannestad:2006nj,Balantekin:2006tg,
Duan:2006jv,Fogli:2007bk,Raffelt:2007cb,Raffelt:2007xt,EstebanPretel:2007yq,
Gava:2008rp,EstebanPretel:2007ec,Duan:2007sh,Duan:2008za,Dasgupta:2007ws,EstebanPretel:2008ni,Gava:2009pj,Galais:2009wi}.
Actually few papers have shown the importance of the one-loop
correction in addition with the neutrino-neutrino interaction.
First it has been shown that for early time after post-bounce when
the matter density profile is very high, neutrinos can encounter
the $\mu-\tau$ resonance and possibly modify the $\nu_e$ flux
\cite{EstebanPretel:2007yq,Duan:2008za}. Unfortunately, it has
also been shown that in such case the high density in addition
with a multi-angle neutrino-neutrino interaction would make those
effects vanish \cite{EstebanPretel:2008ni}. Besides those works,
only one paper\footnote{In \cite{Duan:2008za}, it was noted that
the CP-violating phase could influence sensibly the $\mu$ and
$\tau$ neutrino flavor.} \cite{Gava:2008rp} has shown the
importance of $V_{\mu \tau}$ via the influence of the CP-violating
phase $\delta$ contained in the MNSP matrix and whose value is
still unknown. Such term induces effects of a non-zero
CP-violating phase on the electron neutrino fluxes inside and
outside the supernova.

The goal of this paper is to calculate such corrections in the
SUSY framework with and without taking into account R-parity
breaking interactions. It is organized as follows: Sec.2
introduces the theoretical framework where we briefly remember how
to calculate radiative corrections in this context and more
importantly where we introduce the supersymmetric framework. The
corrections with R-parity conservation are calculated in Sec.3 and
Sec.4 is dedicated to the case where the R-parity is broken.
Before concluding, we explicitly demonstrate in Sec.5 the
influence of $V_{\mu \tau}$ on the $\nu_e$ flux when the
CP-violating phase is non-zero.

\section{Theoretical framework}

\subsection{The calculations of the radiative corrections}

Neutrinos interaction through matter can be described using
indices of refraction and in this case, the evolution equation of
neutrinos with matter, omitting the neutrino-neutrino interaction
is:
\begin{equation}i{d\over dt}\left(\begin{array}{r}\nu_e\\ \nu_\mu\\
\nu_\tau\end{array}\right)=
\left[ {1\over 2p_\nu}U\left(\begin{array}{ccc}\Delta m^2_{12}&0&0\\0&0&0\\
0&0&\Delta m^2_{32}\end{array}\right) U^\dagger
-p_\nu\left(\begin{array}{ccc} \Delta n_{e
\mu}&0&0\\0&0&0\\0&0&\Delta n_{\tau\mu}\end{array}\right)\right]
\left(\begin{array}{r}\nu_e\\ \nu_\mu\\
\nu_\tau\end{array}\right),
\end{equation}
where $U$ is the MNSP matrix, $p_{\nu}$ the neutrino momentum,
$\Delta m^2_{ij}\equiv m_{\nu_i}^2-m_{\nu_j}^2$ and $\Delta
n_{\alpha\beta}\equiv n_{\nu_\alpha}-n_{\nu_\beta}$.

To size the effect of such radiative correction on the neutrino
propagation, we study the one-loop effect on the scattering
amplitude matrix which describes the interaction of neutrinos with
matter:
\begin{equation}\label{SMscatteringamplitude}
M(\nu_\ell f\to \nu_\ell f)=-i{G_F\over
\sqrt{2}}\bar\nu_\ell\gamma^\rho(1-\gamma_5)\nu_\ell\bar
f\gamma_\rho (C^V_{\nu_\ell f}+C^A_{\nu_\ell f}\gamma_5)f .
\end{equation}
In the calculations we will use the same approximations as in
\cite{Botella:1986wy} and \cite{Roulet:1995ef} i.e, neutrinos are
propagating though an unpolarized medium at rest. Consequently,
the neutrino index of refraction can be written as:
\begin{equation}
p_\nu(n_{\nu_\ell}-1)=-\sqrt{2}G_F\sum_{f=u,d,e}C^V_{\nu_\ell
f}N_f .\end{equation} where $N_f$ is the number density of fermion
$f$ in the medium. Therefore, the interesting parameter to study
in order to size the radiative correction to matter interaction is
$C^V_{\nu_\ell f}$ which is defined at the tree-level by
\begin{equation}
C^V_{\nu_\ell f}=T_3(f_L)-2Q_f \sin^2\theta_W+\delta_{\ell f} ,
\end{equation}
with $\sin^2\theta_W\equiv 1-m_W^2/m_Z^2 \simeq 0.23$, while $Q_f$
and $T_3(f_L)$ are respectively the electric charge and the third
component of the weak isospin of the fermion $f_L$. To parametrize
the loop corrections to $C^V_{\nu_\ell f}$ Botella et al.
\cite{Botella:1986wy} have defined in some kind of an arbitrary
but convenient way a new $C^V_{\nu_\ell f}$ by
\begin{equation}
C^V_{\nu_\ell f}=\rho^{\nu_\ell f}T_3(f_L)-2Q_f\lambda^{\nu_\ell
f}s_W^2 .
\end{equation}
for $f\neq \ell$ where the $\rho^{\nu_\ell f}$ includes the
f-dependent (box) diagrams contributions. Since the
$\lambda^{\nu_\ell f}$ are chosen to be independent of the $f$, in
a electrically neutral medium, they will not contribute to $\Delta
n_{\tau\mu}$. Consequently, $\Delta n_{\tau\mu}$ will only be
sensitive to $\Delta\rho^f\equiv \rho^{\nu_\tau f}-\rho^{\nu_\mu
f}$. Note that we are only interested about the difference between
$n_{\nu_\mu}$ and $n_{\nu_\tau}$ indices of refraction because the
loop correction to $\Delta n_{e\mu}=-\sqrt{2}G_FN_e/p_\nu$ will be
negligible since electron neutrinos already encounter charged
current interactions with matter. In the SM, the correction have
been calculated and are found to be small :
\begin{equation}
\Delta n_{\tau\mu}=V_{\mu\tau}=\varepsilon V_e \simeq 5.4 \times
10^{-5}\,V_e
\end{equation}
where $V_{\mu\tau}$ is the effective matter potential to tau
neutrinos due to one-loop corrections and $\varepsilon$ the ratio
between the $V_{\mu \tau}$ and $V_e$ which yields the size of the
loop correction in comparison with the charged-current matter
potential for electron neutrinos $V_e$.

\subsection{The supersymmetric framework}

The most significant theoretical issues of the Standard Model (SM)
are the hierarchy problem for the Higgs mass and the
non-unification of the gauge couplings . Supersymmetry allows us
to address these problems and is an attractive candidate for new
physics beyond the Standard Model (SM). In this theory a
supersymmetric particle called LSP (Lightest Supersymmetric
Particle) is a natural candidate for dark matter. Among various
supersymmetric models the most extensively studied is the minimal
supersymmetric model (MSSM).

The MSSM contains the minimum number of fields to describe the
known SM particles and their superpartners. These fields can be
gathered into chiral superfiels and vector superfields. A chiral
superfield $\hat{\Phi}$ is a multiplet which contains a scalar
field ($z$), a fermionic field ($\psi$) and an auxiliary field
($F$):  $\hat{\Phi}$ = ($z$, $\psi$, $F$). A vector superfield
$\hat{V}$ is a multiplet which contains a bosonic field
($v^{\mu}$), a fermionic field ($\lambda$) and an auxiliary field
($D$): $\hat{V}$ = ($v^{\mu}$, $\lambda$, $D$). $F$ and $D$ are
auxiliary fields, they do not have kinetic terms. They are
eliminated by the minimization equations of the Lagragian. To SM
fermionic fields (leptons, quarks), we have supersymmetric scalar
fields (sleptons, squarks) associated. Concerning SM vector bosons
($U(1)_Y$, $SU(2)_L$, $SU(3)_C$ gauge bosons), fermionic fields
called gauginos (bino, winos, gluinos) are associated to. Finally,
we have fermionic fields called higgsinos associated to Higgs
bosons. Higgsinos and gauginos will mix to generate neutralinos
and charginos.

A superfield is a function of spacetime coordinates and so-called
superspace coordinates which appear as anticommuting Grassman
variables \cite{WessBagger}. It can be expanded in terms of its
component fields and these Grassman variables. Products of
superfields can be developed resulting in products of individual
particle and sparticle fields.

We call superpotential all the renormalizable products of chiral
superfields in a given supersymmetric model. The MSSM's
superpotential is:
\begin{equation}
\begin{array}{rcl}
{\it{W}}_{MSSM} & = & h_{t}\hat{Q}.\hat{H}_{u}\hat{T}_{R}^{c}
 -  h_{b}\hat{Q}.\hat{H}_{d}\hat{B}_{R}^{c}
-  h_{\tau}\hat{L}.\hat{H}_{d}\hat{L}_{R}^{c} \\
& + & \mu \hat{H}_{u}.\hat{H}_{d}
\end{array}
\end{equation}
where $\hat{H}_{u},\hat{Q} ...$ are chiral superfields. The
fermionic part of the supersymmetric lagrangian is obtained
through a procedure where the superfields of the superpotential
are expanded in terms of component fields and then the superspace
coordinates of the result are integrated out:
\begin{equation}
\begin{array}{rcl}
 {\cal{L}}_{MSSM} & = & h_{t} \psi_{Q}.H_{u}\psi_{T_{R}^{c}}
-  h_{b} \psi_{Q}.H_{d}\psi_{B_{R}^{c}} -  h_{\tau} \psi_{L}.H_{d}\psi_{L_{R}^{c}} \\
& & +  \mu \psi_{H_{u}}.\psi_{H_{d}} + \ldots (+ h.c)
\end{array}
\end{equation}

In Supersymmetry, we need two Higgs bosons to give masses to the
other particles.

Despite its simplicity, there are two unexplained hierarchies,
within the MSSM:
\begin{itemize}
 \item [-] the so-called $\mu$-problem \cite{KimNilles}. It arises
 from the presence of a mass $\mu$-term for the Higgs fields in the
superpotential. The only two theoretical natural values for this
parameter are either zero or the Planck energy scale. However, we
need $\mu \gtrsim 100GeV$ to satisfy LEP constraints on the
chargino masses and $\mu \lesssim M_{SUSY}$ for a destabilization
of the Higgs potential in order to have non-vanishing v.e.v. for
the scalar Higgs fields ($<H_{u,d}> \neq 0$)
 \item [-] The other hierarchy with an unknown origin is the one
 existing  between the small neutrino masses (smaller than the eV
 scale) and the electroweak symmetry breaking scale ($\sim 100GeV$).
\end{itemize}

In this paper, our framework is the Next-to Minimal Supersymmetric
extension of the Standard Model (NMSSM) \cite{NMSSM}. This model
provides an elegant solution to the $\mu$-problem via the
introduction of a new gauge-singlet superfield $\hat{S}$ that
acquires naturally a v.e.v. $x$ of the order of the supersymmetry
breaking scale, generating an effective $\mu$ parameter ($\lambda
x = \mu_{\rm eff}$) of order of the electroweak scale.
Furthermore, this model explains the second hierarchy by
generating two neutrino masses at tree level through R-parity
breaking \cite{GregAsmaa:2006} as we will see below.

The NMSSM's superpotential is:
\begin{equation}
\begin{array}{rcl}
{\it{W}}_{NMSSM} & = & h_{t}\hat{Q}.\hat{H}_{u}\hat{T}_{R}^{c}
 -  h_{b}\hat{Q}.\hat{H}_{d}\hat{B}_{R}^{c}
-  h_{\tau}\hat{L}.\hat{H}_{d}\hat{L}_{R}^{c} \\
& + & \lambda \hat{S}\hat{H}_{u}.\hat{H}_{d} +
\frac{\kappa}{3}\hat{S}^{3}.
\end{array}
\end{equation}
Then, the fermionic part of the NMSSM lagrangian is:
\begin{equation}
\begin{array}{rcl}
 {\cal{L}}_{NMSSM} & = & h_{t} \psi_{Q}.H_{u}\psi_{T_{R}^{c}}
-  h_{b} \psi_{Q}.H_{d}\psi_{B_{R}^{c}} -  h_{\tau} \psi_{L}.H_{d}\psi_{L_{R}^{c}} \\
& & +  \lambda S\psi_{H_{u}}.\psi_{H_{d}} + \kappa
S\psi_{S}\psi_{S} + \ldots (+ h.c)
\end{array}
\end{equation}

The NMSSM contains the following particles:
\begin{itemize}
 \item [.] Standard Model fermions and their left and right scalar supersymmetric partners (sfermions)
 \item [.] Standard Model gauge bosons
 \item [.] 3 neutral scalar Higgs ($h_1$, $h_2$, $h_3$)
 \item [.] 2 neutral pseudo-scalar Higgs ($a_1$, $a_2$)
 \item [.] 1charged Higgs ($H^{\pm}$)
 \item [.] 2 charginos ($\chi_{1,2}^{\pm}$) originally from mixing
 between charged fermionic superpartners of the gauge bosons
 (gauginos) and charged fermionic superpartners of Higgs
 bosons (higgsinos)
 \item [.] 5 neutralinos ($\chi^{0}_{1\ldots 5}$) of which the
 lighest called lighest supersymmetric particle (LSP) is gererally
 stable and is thence a natural candidate for Dark Matter. They come
 from the mixing between neutral gauginos et neutral higgsinos.
\end{itemize}

The NMSSM phenomenology contains the MSSM phenomenology if we take
$\lambda \rightarrow 0$, $\kappa \rightarrow 0$ by keeping
$\mu_{\rm eff}$ of the order of $M_W$.

Experimental limits on supersymmetric particle masses obviously
show that SUSY has to be broken. Thus, the supersymmetric particle
masses will be different from Standard Model particle masses. The
soft SUSY breaking terms in the NMSSM are:
\begin{itemize}
 \item [a)] mass terms for  scalar particles:
$$m^{2}_{H_{u}}|H_{u}|^{2} + m^{2}_{H_{d}}|H_{d}|^{2}  + m^{2}_{S}|S|^{2} + m^{2}_{Q_3}|Q_3|^{2}
+ m^{2}_{\tilde{t}_R}|T_{R}|^{2} +  m^{2}_{\tilde{b}_R}|B_{R}|^{2}
+ m^{2}_{L_3}|L_3|^{2} +  m^{2}_{\tilde{\tau}_R}|L_{R}|^{2}
+\ldots
$$ where all the fields are scalar fields,
 \item [b)] mass terms for gauginos:
$$\frac{1}{2}M_{1}\lambda_{1}\lambda_{1} + \frac{1}{2}M_{2}\overrightarrow{\lambda_{2}}.\overrightarrow{\lambda_{2}}
+
\frac{1}{2}M_{3}\overrightarrow{\lambda_{3}}.\overrightarrow{\lambda_{3}}$$
 \item [c)] soft terms associated to the superpotential:
$$ \lambda A_{\lambda}S H_{u}.H_{d}  + \frac{\kappa}{3}A_{\kappa}S^{3}
+ h_{t}A_{t} Q.H_{u}T_{R}^{c}  - h_{b}A_{b} Q.H_{d}B_{R}^{c}  -
h_{\tau}A_{\tau} L.H_{d}L_{R}^{c} +\ldots (+ h.c)$$ where all
fields are scalar fields.
\end{itemize}

\section{R-parity conserved SUSY corrections}

In Supersymmetry, one usually considers all sfermions to be
exactly degenerate at the GUT scale, and obtain their low energy
splittings from the renormalization group evolution of the soft
parameters and from terms arising after the electroweak symmetry
breaking. In this way, although squarks become significantly split
from sleptons, the splittings among the masses of different
slepton generations are only due to the small $\tau$-Yukawa
coupling. This usually implies that $m_{\tilde
\tau}^2-m_{\tilde\mu}^2$ is $O(m_\tau^2)$, and hence the radiative
effects on the $\nu_{\mu,\tau}$ indices of refraction are, in this
case, not larger than the SM ones. In the following, we shall
consider in the following a large $\tilde\tau_L$--$\tilde\tau_R$
mixing. This occurs for large values of the effective $\mu$
parameter and tan$\beta$, or for large values of the parameter $A$
of the trilinear soft terms, in which case the splitting can be
$O[m_\tau(A_{\tau} +\mu_{\rm eff} tan\beta$)] as we can see below.

We give here the stau and smuon mass-squared matrices where we
neglect splittings due to $D$-terms among charged and neutral
sleptons or among $\tilde\ell_L$ and $\tilde\ell_R$. The diagonal
terms are respectively the $\tilde\ell_R$ and $\tilde\ell_L$
mass-squared terms.

For tau sleptons:
\begin{eqnarray}
{\cal{M}}^{2}_{\tilde{\tau}} =   \left(\begin{array}{cc}
     m_{\tau}^{2} + m_{\tilde{\tau}_R}^{2} & m_{\tau}(A_{\tau} + \mu_{\rm eff}tan\beta)  \cr
     m_{\tau}(A_{\tau} + \mu_{\rm eff}tan\beta)  & m_{\tau}^{2} + m_{L_{3}}^{2}  \end{array}\right) \  ,  \
\end{eqnarray}

 and for muon sleptons:
\begin{eqnarray}
{\cal{M}}^{2}_{\tilde{\mu}} =   \left(\begin{array}{cc}
     m_{\mu}^{2} + m_{\tilde{\mu}_R}^{2} & m_{\mu}(A_{\mu} + \mu_{\rm eff}tan\beta)  \cr
     m_{\mu}(A_{\mu} + \mu_{\rm eff}tan\beta)  & m_{\mu}^{2} + m_{L_{2}}^{2}  \end{array}\right)
\end{eqnarray}
where
\begin{itemize}
 \item [-] $m_{\tilde{\tau}_R}^{2}$, $m_{\tilde{\mu}_R}^{2}$, $m_{L_{3}}^{2}$ and
$m_{L_{2}}^{2}$ are of the order of $M_{susy}^{2} \sim
0.1-1TeV^{2}$,
 \item [-] ($A_{\tau} + \mu_{\rm eff}tan\beta$), ($A_{\mu} +
\mu_{\rm eff}tan\beta$) are of the order of $M_{susy}$,
 \item [-] $m_{\tau} = 1.8 GeV$, $m_{\mu} = 105 MeV$.
\end{itemize}

In the $2^{nd}$ generation case, we can neglect the mixing between
$\tilde{\mu}_R$ and $\tilde{\mu}_L$ because the off-diagonal terms
are negligible w.r.t. the diagonal terms. But in the third
generation, we have to consider $\tilde{\tau}_1$, $\tilde{\tau}_2$
instead of $\tilde{\tau}_R$, $\tilde{\tau}_L$.

The SUSY contribution to $\Delta n_{\tau\mu}$ can then be larger
than the SM one.

In the following, to calculate the loop diagrams contributions we
will use the dimensional regularization method and the vanishing
external legs approximation \cite{Botella:1986wy,Roulet:1995ef}
which turns out to be legitimate because of the low masses and
energy of the fermions with respect to the electro-weak ($M_W$)
and SUSY ($M_{SUSY}$) breaking scales. When one does not use this
approximation, the loop integrals to perform will have several
different mass scales. Such calculations are much more complicated
and need advanced mathematical methods as in \cite{Friot:2005cu}.
In this framework, some specific functions will appear:
\begin{itemize}
 \item [-] For the self-energies and the penguin:
 $$H_0(x,y)=\sqrt{xy}\left[{x{\rm ln}x\over (x-y)(x-1)}+(x\leftrightarrow y)\right],$$
$$G_0(x,y)=\left[{x^2{\rm ln}x\over (x-y)(x-1)}+(x\leftrightarrow y)\right].$$
 \item [-] For the box diagrams:
 $$H'(x,y,z)=\sqrt{xy}\left[{x{\rm ln}x\over (x-y)(x-z)(x-1)}+
{y{\rm ln}y\over (y-x)(y-z)(y-1)} +{z{\rm ln}z\over
(z-x)(z-y)(z-1)}\right],$$
$$G'(x,y,z)={x^2{\rm ln}x\over (x-y)(x-z)(x-1)}+
{y^2{\rm ln}y\over (y-x)(y-z)(y-1)} +{z^2{\rm ln}z\over
(z-x)(z-y)(z-1)}.$$
\end{itemize}

\subsection{Vertices}

Following the notation of \cite{haber,nmssmtools1}, we denote by
$N_{ij}$ by $\chi_i^0  \equiv N_{ij}\tilde{\Psi}_j$ where
$\tilde{\Psi}^{T}=
(\tilde{B},\tilde{W}_3,\tilde{h}_u,\tilde{h}_d,\tilde{s})$, $U$
and $V$ are the $2 \times 2$ matrices required for the
diagonalization of the chargino mass matrix.

\begin{figure}[h]
\vspace{.6cm}
\centerline{\includegraphics[scale=0.35,angle=0]{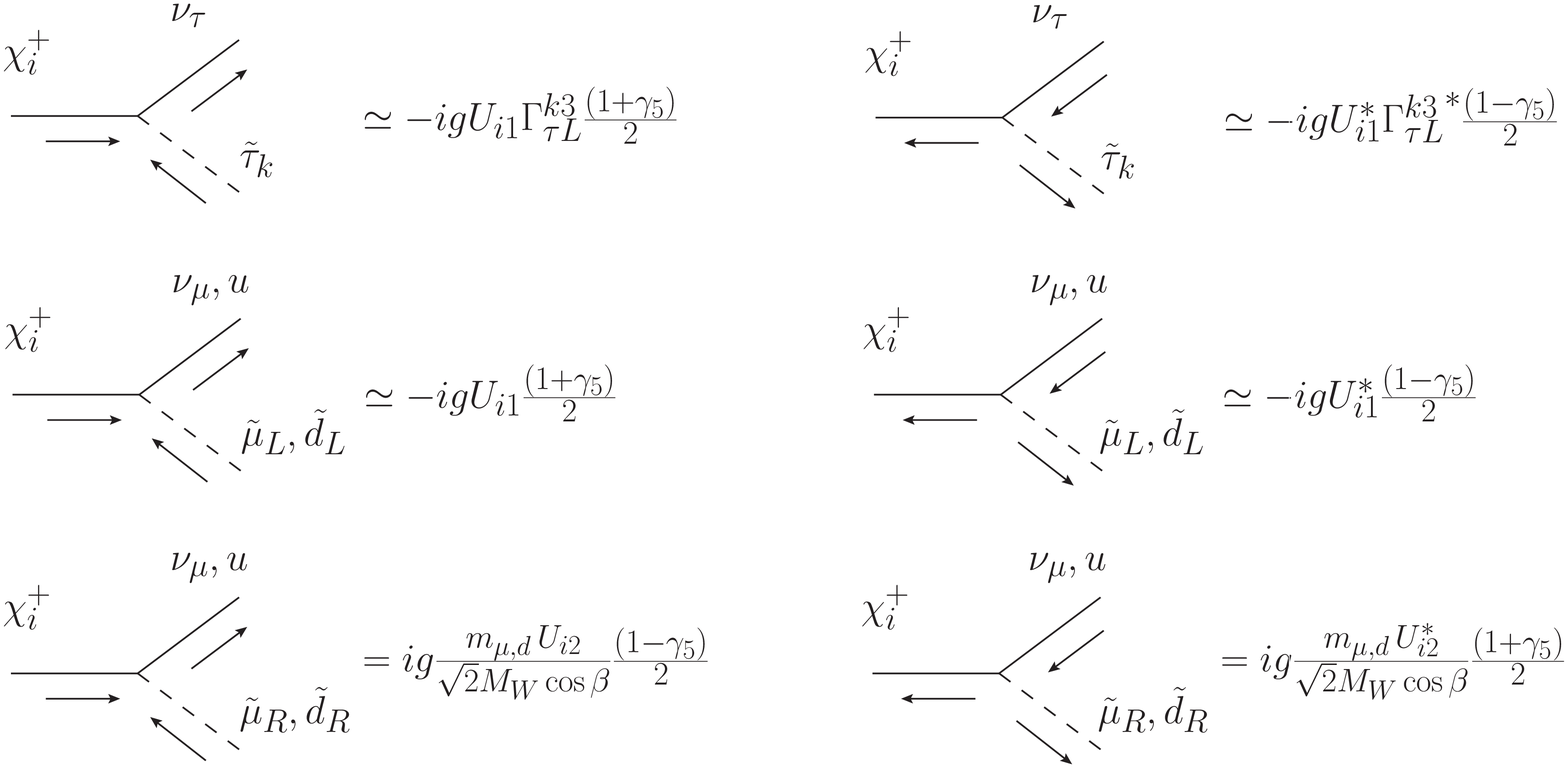}\hspace{.2cm}}
\caption{R-parity conserved vertices involving charginos and up
fermions.} \label{vertex1}
\end{figure}
\begin{figure}[h]
\vspace{.6cm}
\centerline{\includegraphics[scale=0.35,angle=0]{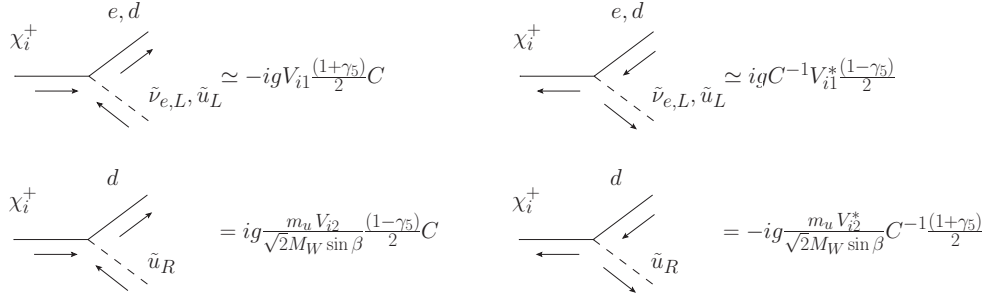}\hspace{.2cm}}
\caption{R-parity conserved vertices involving charginos and down
fermions.} \label{vertex2}
\end{figure}

We see in Fig.(\ref{vertex1}) and (\ref{vertex2}) that vertices
between chargino $\chi_i^+$, a fermion $f$ and the right scalar
partner $\tilde{f}_R$ are negligible w.r.t. vertices between
chargino $\chi_i^+$, a fermion $f$ and the left scalar partner
$\tilde{f}_L$ because $\frac{m_f}{M_W} \ll 1$ for $f \equiv \mu,
u, d$. We will neglect the loops including the primary vertices
below.

\begin{figure}[h]
\vspace{.6cm}
\centerline{\includegraphics[scale=0.35,angle=0]{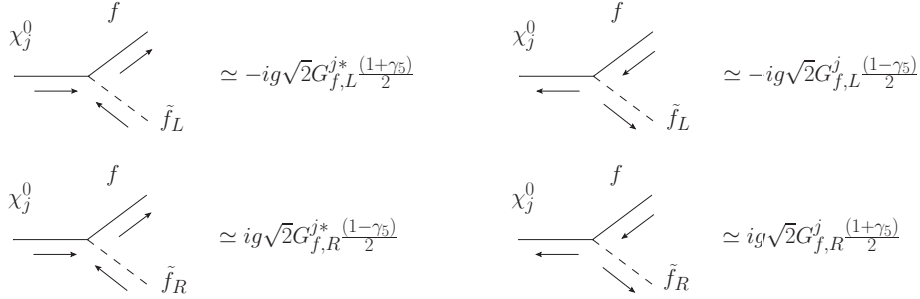}\hspace{.2cm}}
\caption{R-parity conserved vertices involving neutralinos,
fermions and associated sfermions.} \label{vertex3}
\end{figure}
\begin{figure}[h]
\vspace{.6cm}
\centerline{\includegraphics[scale=0.35,angle=0]{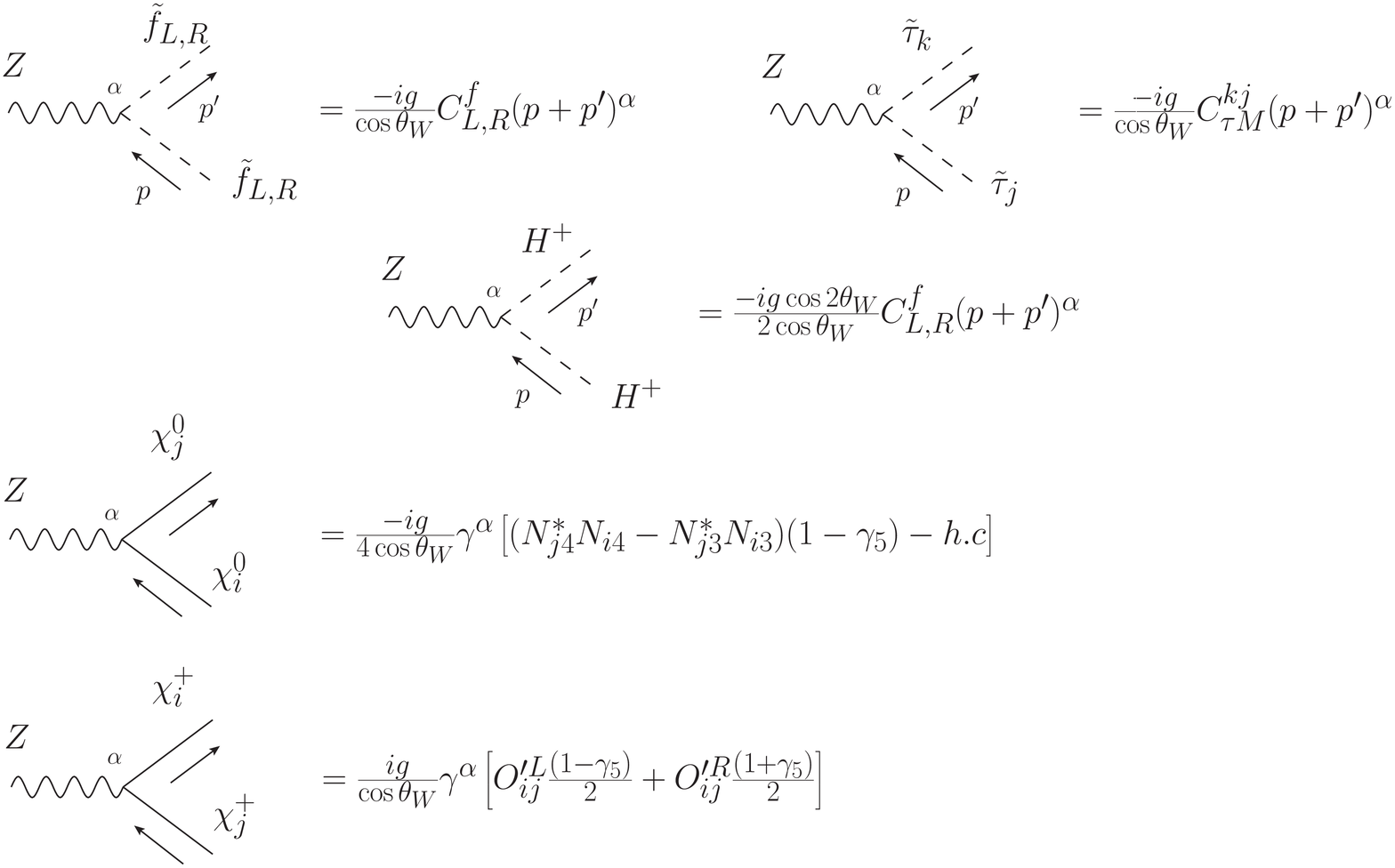}\hspace{.2cm}}
\caption{R-parity conserved vertices with $Z^0$-boson.}
\label{vertex4}
\end{figure}

The coupling of one neutralino $\chi_i^0$ to one fermion and the
associated left (right) scalar partner $G^j_{fL(R)}$ is:
$$G^j_{fL}=Q_fsin\theta_W N^{*}_{j1}+{c_L^f\over cos\theta_W}N^{*}_{j2}$$
where $f \equiv \nu, e, u, d$.
$$G^j_{fR}={\rm sign}(m_{\chi^0_j})\left[ Q_f sin\theta_W N_{j1} + {c_R^f\over
cos\theta_W}N_{j2} \right]$$ where $f \equiv e, u, d$.

The coupling of left (right) scalar partners to the $Z^0$-boson
is:
$$c^f_{L(R)}=T_3(f_{L(R)})-Q_f sin^2\theta_W.$$
The coupling of scalar partners in their mass basis to the
$Z^0$-boson \cite{be91} is:
$$c^{kj}_{f \it{M}}=T_3(f)\sum_{i=1}^{3}\Gamma_{fL}^{ki}\Gamma_{fL}^{ji *}
- Q_f sin^2\theta_W \delta^{kh}$$ where $\tilde{\tau}_L =
\Gamma_{\tau L}^{k3}\tilde\tau_k$.  The coupling of charginos to
the $Z^0$-boson \cite{be91,rosiek}, is:
$${\cal O'}^L_{ij}=-V_{i1}V^*_{j1}-\frac{1}{2}V_{i2}V^*_{j2}+\delta_{ij}sin^2\theta_W$$
and
$${\cal O'}^R_{ij}=-U_{i1}U^*_{j1}-\frac{1}{2}U_{i2}U^*_{j2}+\delta_{ij}sin^2\theta_W.$$

We will assume below that all the parameters are real.

\subsection{Self-energy diagrams}

The computation of the supersymmetric contribution to $\Delta\rho$
requires the evaluation of the Feynman diagrams.

The self-energy, penguin and box contributions to $\Delta\rho^f$
can be written as
$$\Delta\rho^f=\Delta\rho_p+{\Delta\rho^f_{box}\over T_3(f_L)}.$$

We consider here the contributions to neutrino and antineutrino
scattering involving self-energy diagrams.

\begin{figure}[h]
\vspace{.6cm}
\centerline{\includegraphics[scale=0.3,angle=0]{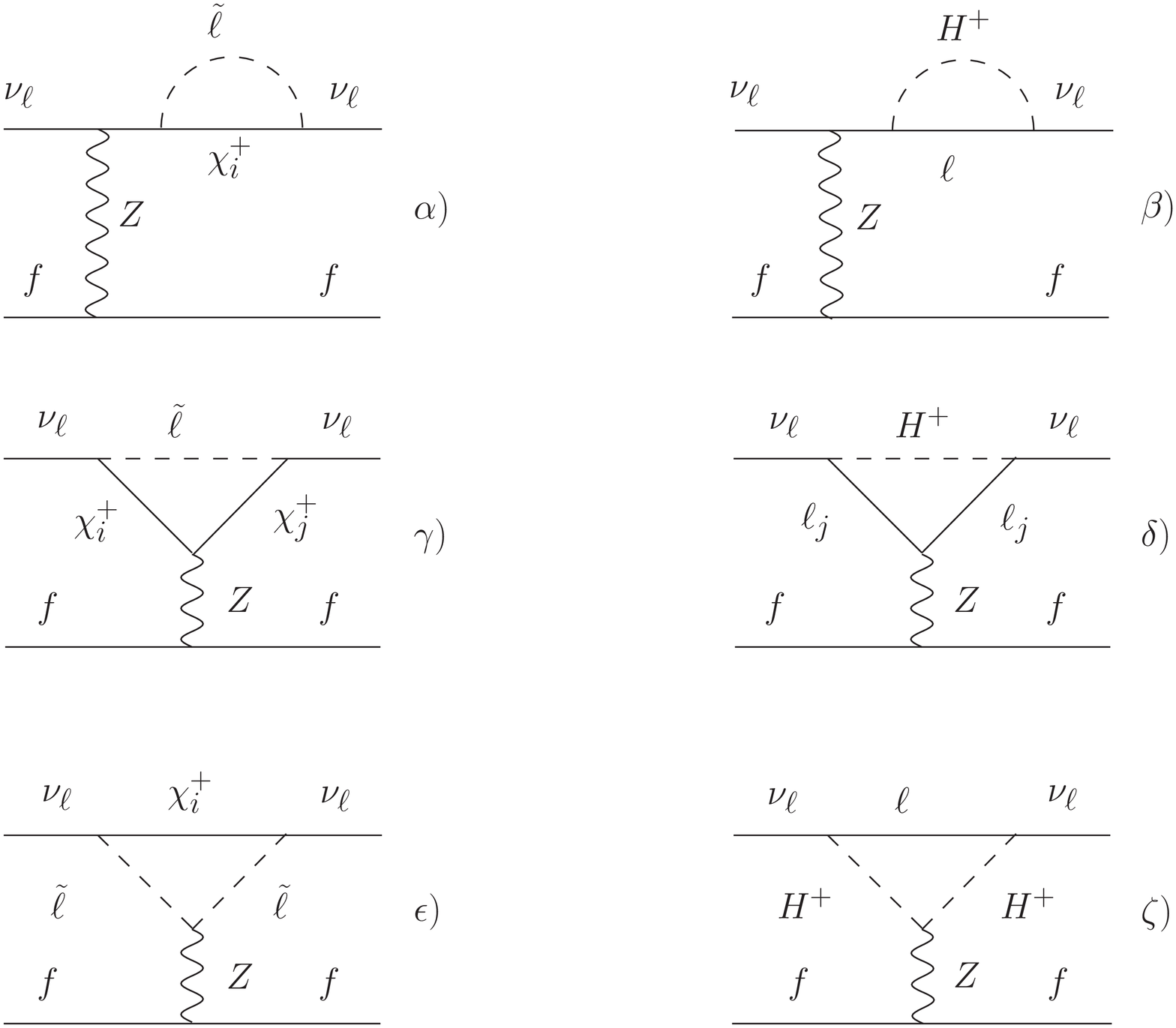}\hspace{.2cm}}
\caption{R-parity conserved penguins and self-energies involving
charginos and the charged Higgs.} \label{selfpengsleptonHcharged}
\end{figure}

The first contribution implies a slepton-chargino loop
(Fig.\ref{selfpengsleptonHcharged}$\alpha$). We diagonalize the
stau mass matrix and use the vertices of Fig.(\ref{vertex1}). We
neglect the $\tilde{\mu}_R$ contribution.

\begin{equation}\label{Self1}
\begin{array}{rcl}
 \Delta\rho^{\tilde \ell}(\Sigma)=-{\alpha_W\over 8\pi} \sum_{j=1}^2
 U_{j1}^2 &&\left[  \sum_{k=1}^2 {\Gamma_{\tau L}^{k3}}^2  \left\{
G_0(X_{\chi^+_j\tilde\tau_k},1)+{\rm ln} {m_{\tilde\tau_k}^2\over
\mu^2} \right\} \right. \\
&& \left. -\left\{ G_0(X_{\chi^+_j\tilde\mu_L},1)+{\rm ln}
{m_{\tilde\mu_L}^2\over \mu^2} \right\} \right]
\end{array}
\end{equation}

where $X_{ab}=\frac{m_a^2}{m_b^2}$. Then, we have a
neutralino-sneutrino loop (Fig.\ref{selfpengneutralino}$\iota$).

\begin{eqnarray}\label{Self2}
 \Delta\rho^{\tilde{\nu}}(\Sigma)=-{\alpha_W\over 4\pi}
\sum_{j=1}^5 {G_{\nu L}^{j}}^2\left\{
G_0(X_{\chi^0_j\tilde\nu_{\tau_L}},1)+{\rm ln}
{m_{\tilde\nu_{\tau_L}}^2\over \mu^2}-(\tilde\nu_{\tau_L}\to
\tilde\nu_{\mu_L}) \right\}.
\end{eqnarray}

\subsection{penguin diagrams}

We consider here the contributions to (anti)neutrino scattering
involving penguins. We use the vertices of Fig.(\ref{vertex4}).

\begin{figure}[h]
\vspace{.6cm}
\centerline{\includegraphics[scale=0.3,angle=0]{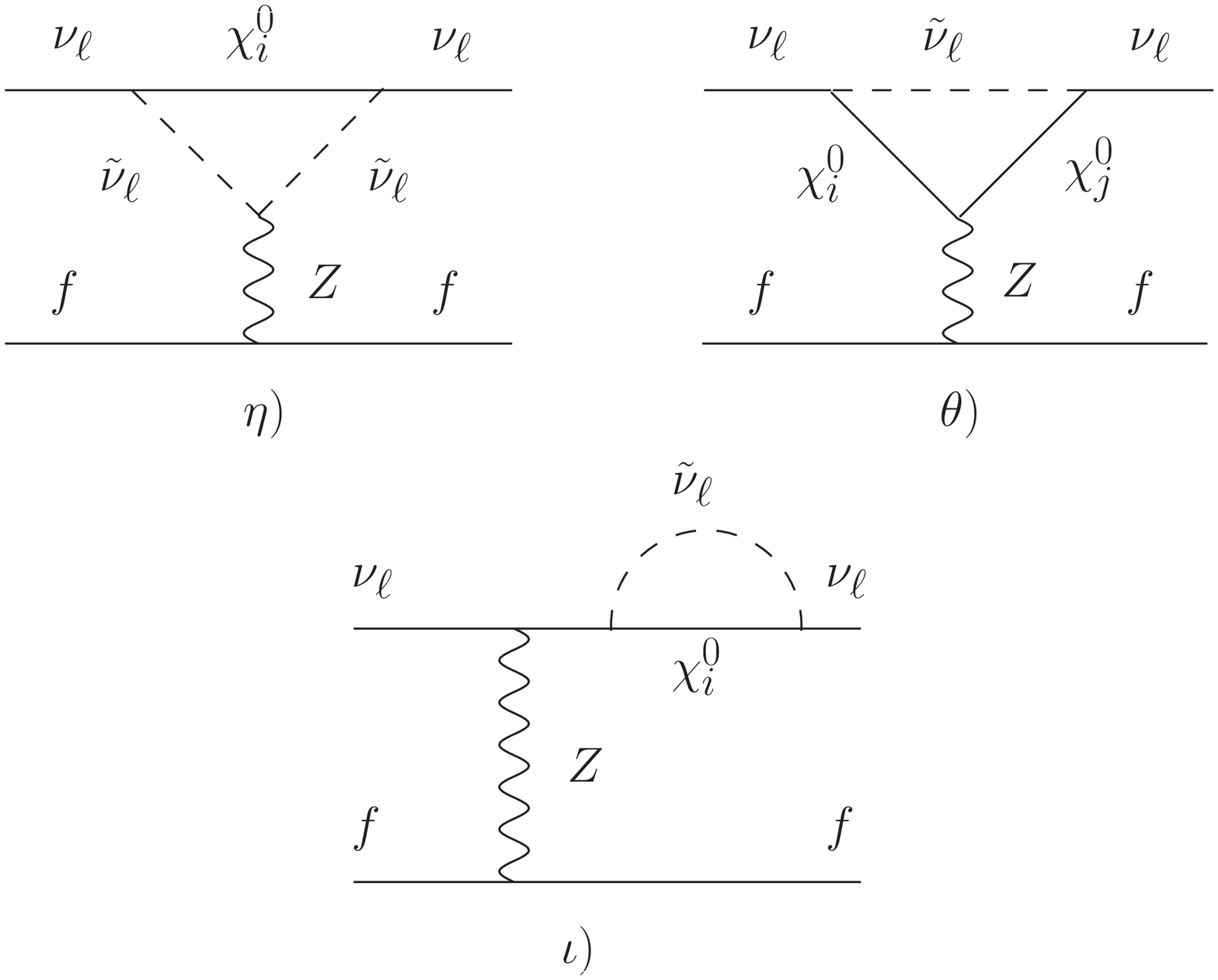}\hspace{.2cm}}
\caption{R-parity conserved penguins and self-energies involving
neutralinos.} \label{selfpengneutralino}
\end{figure}

$\Delta\rho_p(\tilde \ell)$ implies a chargino-slepton loop in
which the slepton couples to the $Z^0$-boson
(Fig.\ref{selfpengsleptonHcharged}$\varepsilon$). In
$\Delta\rho_p(\chi^+)$, the chargino couples to the $Z^0$-boson
(Fig.\ref{selfpengsleptonHcharged}$\gamma$).
$\Delta\rho^L_p(\tilde \nu)$ implies a neutralino-sneutrino loop
in which the sneutrino couples to the $Z^0$-boson
(Fig.\ref{selfpengneutralino}$\eta$) and in
$\Delta\rho^L_p(\chi^0)$ the neutralino couples to $Z^0$-boson
(Fig.\ref{selfpengneutralino}$\theta$).

\begin{equation}\label{Peng1}
\begin{array}{rcl}
 \Delta\rho_p(\tilde \ell)={\alpha_W\over 4\pi }
\sum_{i=1}^2 U_{i1}^2 &&\left[  \sum_{j,k=1}^2 \Gamma_{\tau
L}^{k3}\Gamma_{\tau L}^{j3}c^{kj}_{\tau}  \left\{
G_0(X_{\tilde\tau_j \chi^+_i},X_{\tilde\tau_k \chi^+_i})+{\rm ln}
{m_{\chi^+_i}^2\over
\mu^2} \right\} \right. \\
&& \left. - c^\ell_L \left\{ G_0(X_{\chi^+_i\tilde\mu_L},1)+{\rm
ln} {m_{\tilde\mu_L}^2\over \mu^2} \right\} \right],
\end{array}
\end{equation}

\begin{eqnarray}\label{Peng2}
\Delta\rho^L_p(\tilde \nu)={\alpha_W\over 4\pi } \sum_{j=1}^5
{G_{\nu L}^{j}}^2 \left\{
G_0(X_{\chi^0_j\tilde\nu_{\tau_L}},1)+{\rm ln}
{m_{\tilde\nu_{\tau_L}}^2\over \mu^2}-(\tilde\nu_{\tau_L}\to
\tilde\nu_{\mu_L}) \right\},
\end{eqnarray}

\begin{equation}\label{Peng3}
\begin{array}{rcl}
\Delta\rho_p(\chi^+) &=& {\alpha_W\over 4\pi }
\sum_{i,j=1}^2U_{i1}U_{j1} \times \\
&& \left[ \sum_{k=1}^2 {\Gamma_{\tau L}^{k3}}^2 \left\{2{\cal
O'}^L_{ij} H_0(X_{\chi^+_i\tilde\tau_k},X_{\chi^+_j\tilde\tau_k}
)- {\cal O'}^R_{ij}
\left(G_0(X_{\chi^+_i\tilde\tau_k},X_{\chi^+_j\tilde\tau_k})+{\rm
ln}{m_{\tilde\tau_k}^2\over\mu^2}\right) \right\}\right.\\
&- &\left.\left\{2{\cal O'}^L_{ij}
H_0(X_{\chi^+_i\tilde\mu_L},X_{\chi^+_j\tilde\mu_L} )- {\cal
O'}^R_{ij}
\left(G_0(X_{\chi^+_i\tilde\mu_L},X_{\chi^+_j\tilde\mu_L})+{\rm
ln}{m_{\tilde\mu_L}^2\over\mu^2}\right) \right\} \right]
,\end{array}
\end{equation}

\begin{equation}\label{Peng4}
\begin{array}{rcl}
\Delta\rho^L_p(\chi^0)&=&{\alpha_W\over 4\pi }
\sum_{i,j=1}^2G_{\nu L}^{i}G_{\nu L}^{j} \left\{N_{i4}N_{j4} -
N_{i3}N_{j3} \right\} \times \\
&&
\left[2H_0(X_{\chi^0_i\tilde\nu_{\tau_L}},X_{\chi^0_j\tilde\nu_{\tau_L}}
) +
\left\{G_0(X_{\chi^0_i\tilde\nu_{\tau_L}},X_{\chi^0_j\tilde\nu_{\tau_L}})+{\rm
ln}{m_{\tilde\nu_{\tau_L}}^2\over\mu^2}\right\}-(\tilde\nu_{\tau_L}\to
\tilde\nu_{\mu_L}) \right].
\end{array}
\end{equation}

We have neglected the loops involving $\tilde{\mu}_R$.

When we sum all the contributions
(Fig.\ref{selfpengsleptonHcharged}$\beta$,
\ref{selfpengsleptonHcharged}$\delta$,
\ref{selfpengsleptonHcharged}$\zeta$) involving the charged Higgs
boson \cite{Roulet:1995ef}, we find:
$$\Delta\rho_p^{H^+}\simeq -{\alpha_W\over 4\pi}{m_\tau^2\over
M_W^2}{\rm tg}^2\beta{y\over 2}\left[{1\over 1-y}+{{\rm ln}y\over
(1-y)^2}\right].$$

\subsection{Box diagrams}

\begin{figure}[h]
\vspace{.6cm}
\centerline{\includegraphics[scale=0.3,angle=0]{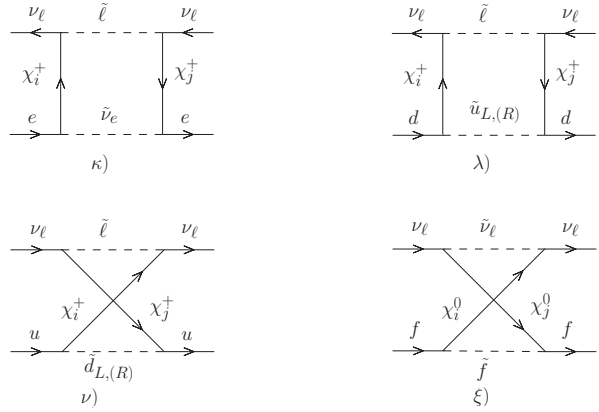}\hspace{.2cm}}
\caption{R-parity conserved box diagrams.} \label{boxRPC}
\end{figure}

The charge conjugation operators in the vertices of
Fig.(\ref{vertex2}) imply that all box diagrams involving
charginos induce radiative corrections to neutrino scattering
only, despite the fact that the neutrino fermionic lines of
Fig.(\ref{boxRPC}$\kappa$) and Fif.(\ref{boxRPC}$\lambda$) are
oriented on the left. As we did before, we diagonalize the stau
mass matrix and we neglect the $\tilde{\mu}_R$ contribution. The
box diagrams involving charginos (resp. Fig.\ref{boxRPC}$\kappa$,
\ref{boxRPC}$\lambda$, \ref{boxRPC}$\nu$) are:
\begin{equation}\label{Box1}
\begin{array}{rcl}
\Delta\rho^e_{box}(\chi^+)&=&-{\alpha_W\over 4\pi }{M_W^2\over
m_{\tilde\nu_e}^2}\sum_{j,k=1}^2V_{j1}V_{k1}U_{j1}U_{k1} \times \\
&\ & \left\{\sum_{i=1}^2 {\Gamma_{\tau L}^{i3}}^2
H'(X_{\chi^+_j\tilde\nu_e}, X_{\chi^+_k \tilde\nu_e},
X_{\tilde\tau_i\tilde\nu_e})-H'(X_{\chi^+_j\tilde\nu_e},
X_{\chi^+_k \tilde\nu_e}, X_{\tilde\mu_L\tilde\nu_e})\right\},
\end{array}
\end{equation}

\begin{equation}\label{Box2}
\begin{array}{rcl}
\Delta\rho^d_{box}(\chi^+)=\Delta\rho^e_{box}(\chi^+)\
(\tilde\nu_e\to\tilde u_L),
\end{array}
\end{equation}

\begin{equation}\label{Box3}
\begin{array}{rcl}
\Delta\rho^u_{box}(\chi^+)= -{\alpha_W\over 8\pi }{M_W^2\over
m_{\tilde d_L}^2}\sum_{j,k=1}^2  U_{j1}^2 U_{k1}^2
&&\left\{\sum_{i=1}^2 {\Gamma_{\tau L}^{i3}}^2
G'(X_{\tilde\tau_i\tilde d_L},X_{\chi^+_j\tilde d_L}, X_{\chi^+_k
\tilde d_L}) \right.\\
&-&\left. G'(X_{\tilde\mu_L\tilde d_L},X_{\chi^+_j\tilde d_L},
X_{\chi^+_k \tilde d_L})\right\}.
\end{array}
\end{equation}

The box diagrams involving neutralinos with crossed fermionic
lines will only contribute to neurino scattering because the
neutrino fermionic lines are oriented on the right. Here,
$\tilde{f}_R$ contributions are non-negligible.

We consider here the box diagrams involving neutralinos for
neutrino scattering (Fig.\ref{boxRPC}$\xi$):
\begin{itemize}
  \item [.] with left sfermions ($\tilde{e}_L$, $\tilde{u}_L$,
  $\tilde{d}_L$):
\begin{equation}\label{Box4}
\begin{array}{rcl}
\Delta\rho^{\tilde{f}_L}_{box}(\chi^0)&=& -{\alpha_W\over 2\pi
}\sum_{j,k=1}^5 G^k_{\nu L}G^{j}_{\nu
L}G^k_{fL}G^{j}_{fL}{M_W^2\over m_{\tilde f_L}^2}\left[
G'(X_{\tilde\nu_\tau\tilde f_L}, X_{\chi^0_j\tilde f_L},
X_{\chi^0_k\tilde f_L})-(\tilde\nu_\tau\to \tilde\nu_\mu)\right]
\end{array}
\end{equation}
  \item [.] with right sfermions ($\tilde{e}_R$, $\tilde{u}_R$,
  $\tilde{d}_R$):
\begin{equation}\label{Box5}
\begin{array}{rcl}
\Delta\rho^{\tilde{f}_R}_{box}(\chi^0)&=& {\alpha_W\over \pi
}\sum_{j,k=1}^5 G^k_{\nu L}G^{j}_{\nu
L}G^k_{fR}G^{j}_{fR}{M_W^2\over m_{\tilde f_R}^2}\left[
H'(X_{\chi^0_j\tilde f_R}, X_{\chi^0_k\tilde f_R},
X_{\tilde\nu_\tau\tilde f_R}) -(\tilde\nu_\tau\to
\tilde\nu_\mu)\right]
\end{array}
\end{equation}
\end{itemize}

Note that the contributions from all box diagrams involving
antineutrino scattering are identical to the previous
contributions, only the forms of the boxes will be different. For
instance if a ladder box was contributing to antineutrinos then
the corresponding box for neutrinos will have crossed fermionic
lines and vice-versa.

%\begin{equation}\label{Box3}
%\begin{array}{rcl}
%\Delta\rho^f_{box}(\chi^0)&=& -{\alpha_W\over 2\pi }\sum_{j,k=1}^5
%G^k_{\nu L}G^{j}_{\nu L}G^k_{fL}G^{j}_{fL}{M_W^2\over m_{\tilde
%f_L}^2}\left[ G'(X_{\tilde\nu_\tau\tilde f_L}, X_{\chi^0_j\tilde
%f_L}, X_{\chi^0_k\tilde f_L})-(\tilde\nu_\tau\to
%\tilde\nu_\mu)\right]
%\end{array}
%\end{equation}

%\begin{equation}\label{Box4}
%\begin{array}{rcl}
%\Delta\rho^f_{box}(\chi^0)&=& {\alpha_W\over \pi }\sum_{j,k=1}^5
%G^k_{\nu L}G^{j}_{\nu L}G^k_{fR}G^{j}_{fR}{M_W^2\over m_{\tilde
%f_R}^2}\left[ H'(X_{\chi^0_j\tilde f_R}, X_{\chi^0_k\tilde f_R},
%X_{\tilde\nu_\tau\tilde f_R}) -(\tilde\nu_\tau\to
%\tilde\nu_\mu)\right]
%\end{array}
%\end{equation}

\subsection{Numerical results}

We make here a low-energy study where we take first generation
sleptons degenerate with the second generation ones, and only
allow the third generation sleptons to have a different mass. It
is a possibility that sfermion masses may dynamically align along
the directions, in flavour space, of the fermion masses,
suppressing FCNC but allowing large mass splittings \cite{di95}.

\begin{figure}[h]
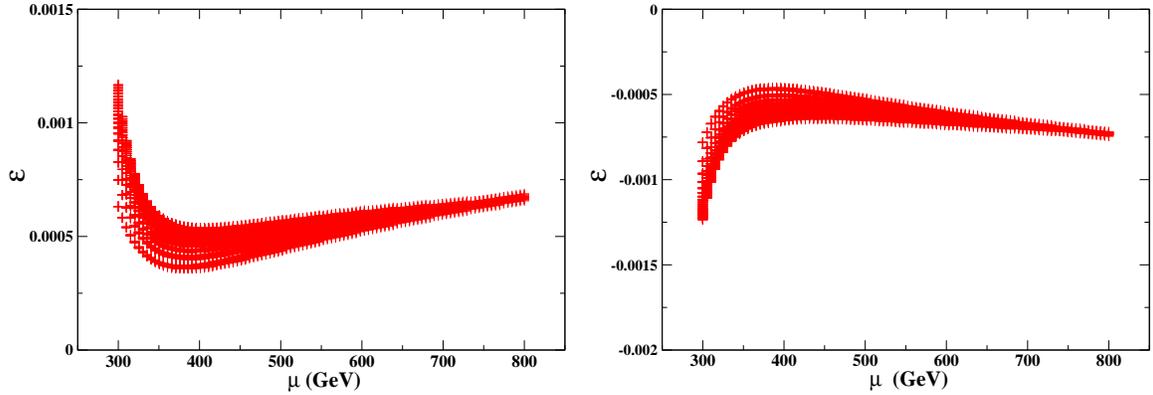

\vspace{.6cm}
\centerline{\includegraphics[scale=0.3,angle=0]{plot1an.eps}\hspace{.2cm}
\includegraphics[scale=0.3,angle=0]{plot1bn.eps}}
\caption{$\varepsilon$ as a function of $\mu$. We fix
$\lambda=0.4$ and $\kappa=0.5$. $M_1 = 66GeV$, $M_2 = 133GeV$ and
$M_3 = 500GeV$. The figure on the left represents the normal
supersymmetric hierarchy ($m_{\tilde\tau}=300 GeV$ and
$m_{\tilde\mu}=200 GeV$), the figure on the right represents the
inverted supersymmetric hierarchy ($m_{\tilde\tau}=200 GeV$ and
$m_{\tilde\mu}=300 GeV$).} \label{plot1}
\end{figure}

We consider two experimentally allowed cases for the sleptons with
a splitting between the third and the second generation:
\begin{itemize}
 \item [.] $m_{\tilde\tau}=300 GeV$ and $m_{\tilde\mu}=200 GeV$
 (normal supersymmetric hierarchy)
 \item [.] $m_{\tilde\tau}=200 GeV$ and $m_{\tilde\mu}=300 GeV$
 (inverted supersymmetric hierarchy)
\end{itemize}

We also assume that squarks are much heavier than sleptons:
$M_{\tilde{Q}}=1TeV$. These are the effects of gluino masses in
the renormalization group evolution of scalar masses.

The splitting among the sleptons of the second and third
generations is mainly responsible for the size of $\Delta\rho$.

\begin{figure}[h]
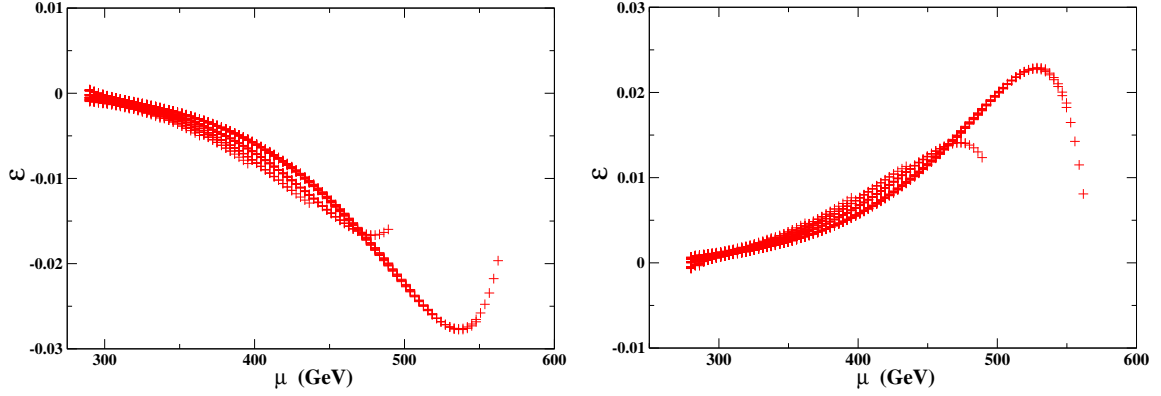

\vspace{.6cm}
\centerline{\includegraphics[scale=0.3,angle=0]{plot2an.eps}\hspace{.2cm}
\includegraphics[scale=0.3,angle=0]{plot2bn.eps}}
\caption{$\varepsilon$ as a function of $\mu$. We fix
$\lambda=0.4$ and $\kappa=0.5$. $M_1 = 150GeV$, $M_2 = 300GeV$ and
$M_3 = 1TeV$. The figure on the left represents the normal
supersymmetric hierarchy, the figure on the right represents the
inverted supersymmetric hierarchy.} \label{plot2}
\end{figure}

The purpose of this Section is the investigation of the
supersymmetric parameter space in some specific cases. To this end
we made a subroutine to the Fortran code NMHDECAY, which is
available on the NMSSMTools web page \cite{nmssmtools1},
\cite{nmssmtools2}, \cite{nmssmtools3}. This subroutine computes
the different R-parity conserved supersymmetric contributions to
$\varepsilon$. Note that $\varepsilon$ is the same for
antineutrinos, as in the SM.

By making scans over the supersymmetric parameter space, we can
obtain, in some regions of the parameter space, divergences for
either $m_{\chi_j^+}=m_{\tilde{\tau}_k}$,
$m_{\chi_j^+}=m_{\tilde{\mu}_L}$ or
$m_{\chi_i^0}=m_{\tilde{\nu}_\ell}$. This is because we assume
vanishing external legs.

In supersymmetry, there are many parameters. $\varepsilon$ doesn't
depend very much on $\lambda$ and $\kappa$ so we fix them as we
usually do in supersymmetry: $\lambda=0.4$ and $\kappa=0.5$ . We
allow some other parameters to vary: ${\rm tan}\beta$, $M_A$ and
$\mu_{\rm eff}$.

${\rm tan}\beta$ is the ratio $v_u / v_d$ where $v_u$ and $v_d$
are the v.e.v. of the scalar Higgs fields $H_u$ and $H_d$. $M_A$
is an effective supersymmetric parameter somewhat equivalent to
the second pseudoscalar Higgs mass in our regions of the parameter
space.\\ \\ \\ \\ We show two different illustrative situations
motivated by high-energy models for the gaugino masses:
\begin{itemize}
 \item [.] $M_1 = 66GeV$, $M_2 = 133GeV$ and $M_3 = 500GeV$
 (cf Fig.(\ref{plot1}))
 \item [.] $M_1 = 150GeV$, $M_2 = 300GeV$ and $M_3 = 1TeV$
 (cf Fig.(\ref{plot2}))
\end{itemize}

In Fig.(\ref{plot1}), ${\rm tan}\beta$ varies between 2 and 15,
$M_A$ varies between 579 and 2000 GeV, $\mu$ varies between 300
and 800 GeV. We see in Fig.(\ref{plot1}) that contrary to the SM
case, $V_{\mu\tau}$ can be either positive or negative.

In Fig.(\ref{plot2}), ${\rm tan}\beta$ varies between 2 and 18
(for the figure on the left) and between 2 and 9.6 (for the figure
on the right), $M_A$ varies between 500 and 1000 GeV, $\mu$ varies
between 200 and 564 GeV.

On the other hand, as we can see in the Fig.(\ref{plot2}),
$\varepsilon$ can go up to $2\times \,10^{-2}$ in this region of
the parameter space. The sign of $\mu$ does not have an important
impact on the maximal value that $\varepsilon$ can reach.

%\begin{figure}[h]
%\vspace{.6cm}
%\centerline{\includegraphics[scale=3,angle=0]{plot1_-mu(final)_hi_2-3.jpg}\hspace{.2cm}
%\includegraphics[scale=3,angle=0]{plot1_-mu(final)_hn_3-2.jpg}}
%\caption{} \label{plot1_-mu(final)}
%\end{figure}

We finally mention that other extensions of the SM may also lead
to sizable effects upon $\Delta n_{\tau\mu}$. In the next section,
we will consider supersymmetric $R$-parity violating interactions
which can have important effects on the neutrino indices of
refraction already at the tree-level~\cite{ro91}.

\section{R-parity breaking SUSY corrections}

\subsection{Introduction}

The superpotential $\it{W}_{NMSSM}$ is not the most general
superpotential we wan write because there exist some other gauge
invariant couplings that we didn't take into account. These new
couplings break a discrete symmetry called R-parity. This symmetry
requires that an interaction must have an even number of SUSY
particles. If R-parity is broken, the LSP will no longer be stable
\cite{CCGreg}. R-parity violating interactions also violate lepton
number ($\it{L}$) or baryon number ($\it{B}$). To be as general as
possible, the superpotential has to contain the following terms:

%$\slashed{R}_P$

%\begin{eqnarray}\label{WRpv}
% \it{W}_{\displaystyle{\not\?}R_{P}} = \sum_{i,j,k}\left(\frac{1}{2}\lambda_{ijk}\hat{L}_{i}\hat{L}_{j}\hat{E}_{k}^{c}
% + \lambda'_{ijk}\hat{L}_{i}\hat{Q}_{j}\hat{D}_{k}^{c}
% + \frac{1}{2}\lambda''_{ijk}\hat{U}_{i}^{c}\hat{D}_{j}^{c}\hat{D}_{k}^{c}
% + \mu_{i}\hat{H}_{u}\hat{L}_{i} + \lambda_{i}\hat{S}\hat{H}_{u}\hat{L}_{i} \right)
%\end{eqnarray}

\begin{eqnarray}\label{WRpv}
 \it{W}_{\slashed{R}_P} = \sum_{i,j,k}\left(\frac{1}{2}\lambda_{ijk}\hat{L}_{i}\hat{L}_{j}\hat{E}_{k}^{c}
 + \lambda'_{ijk}\hat{L}_{i}\hat{Q}_{j}\hat{D}_{k}^{c}
 + \frac{1}{2}\lambda''_{ijk}\varepsilon_{\alpha\beta\gamma}\hat{U}_{i}^{c\alpha}\hat{D}_{j}^{c\beta}\hat{D}_{k}^{c\gamma}
 + \mu_{i}\hat{H}_{u}\hat{L}_{i} + \lambda_{i}\hat{S}\hat{H}_{u}\hat{L}_{i} \right)
\end{eqnarray}

The lagrangian ${\cal{L}}_{\slashed{R}_P}$ can be derived using
the common procedure \cite{Chemtob,GregAsmaa:2006}.

If we consider this part of the superpotential:
\begin{eqnarray}\label{superpotentielmassesRPV}
 \it{W}_{masses} = \it{W}_{NMSSM}
 + \mu_{i}\hat{H}_{u}\hat{L}_{i} + \lambda_{i}\hat{S}\hat{H}_{u}\hat{L}_{i}
\end{eqnarray}

we can generate two neutrino masses at tree-level
\cite{GregAsmaa:2006} by mixing neutrinos and neutralinos. By
considering a See-Saw-like mechanism, the mass matrix is:

\begin{eqnarray}
{\cal{M}}_{\tilde{\chi^{0}}} =   \left(\begin{array}{cc}
{\cal{M}}_{NMSSM}   &   \xi_{\slashed{R}_P}^{T} \cr
\xi_{\slashed{R}_P}  &  0_{3\times 3}
\end{array}\right)
\end{eqnarray}

where ${\cal{M}}_{NMSSM}$ is the R-parity conserved neutralino
mass matrix and $\xi_{\slashed{R}_P}$ is the part of the mass
matrix induced by R-parity violation which mix neutralinos and
neutrinos. We will assume here: $v_i/v_{u,d} \ll 1$, $|\mu_i/\mu|
\ll 1$ and $|\lambda_i/\lambda| \ll 1$ in order to reproduce the
neutrino phenomenology. $v_i$ are the v.e.v. of the sneutrinos.
\\
This model is self-consistent because we give here a way to
generate the neutrino masses contrary to many other scenarios of
radiative corrections on neutrino indices of refraction.

%\begin{eqnarray}
%{\cal{M}}_{NMSSM} =   \left(\begin{array}{ccccc}
%     M_{1}      &        0          &  \frac{g_{1}v_{u}}{\sqrt{2}} & -\frac{g_{1}v_{d}}{\sqrt{2}} &                         0                         \cr
%                &     M_{2}         & -\frac{g_{2}v_{u}}{\sqrt{2}} &  \frac{g_{2}v_{d}}{\sqrt{2}} &                         0                         \cr
%                &                   &               0              &          - \lambda x         &  -\lambda v_{d} + \sum_{i=1}^{3}\lambda_{i}v_{i}  \cr
%                &                   &                              &               0              &                  -\lambda v_{u}                   \cr
%                &                   &                              &                              &                     2\kappa x                     \end{array}\right)
%\end{eqnarray}

%\begin{eqnarray}
%\xi_{\slashed{R}_P} =   \left(\begin{array}{ccccc}
%-\frac{g_{1}v_{1}}{\sqrt{2}} & \frac{g_{2}v_{1}}{\sqrt{2}} &
%\mu_{1}+\lambda_{1}x & 0 & \lambda_{1}v_{u} \cr
%-\frac{g_{1}v_{2}}{\sqrt{2}} & \frac{g_{2}v_{2}}{\sqrt{2}} &
%\mu_{2}+\lambda_{2}x & 0 & \lambda_{2}v_{u} \cr
%-\frac{g_{1}v_{3}}{\sqrt{2}} & \frac{g_{2}v_{3}}{\sqrt{2}} &
%\mu_{3}+\lambda_{3}x & 0 & \lambda_{3}v_{u}  \end{array}\right)
%\end{eqnarray}

In the following, $\nu_i = N_{ij}\tilde{\Psi}_j$ where
$\tilde{\Psi}^{T} =
(\tilde{B},\tilde{W}_3,\tilde{h}_u,\tilde{h}_d,\tilde{s},\nu_e,\nu_{\mu},\nu_{\tau})$,
$i=1..3$ and $j=1..8$. $\nu_e$, $\nu_{\mu}$ and $\nu_{\tau}$
represent the neutrinos in their flavour basis, $\nu_i$ represent
the neutrinos in their mass basis. $\chi^0_i =
N_{(i+3)j}\tilde{\Psi}_j$ where $i=1..5$ and $j=1..8$. $\chi^0_i$
are the five NMSSM's neutralinos.

\subsection{Tree-level corrections}
 Because of R-parity breaking, new interactions are possible between SUSY particles. This yields new
one-loop corrections but also tree level corrections contrary to
the R-parity conserved case. Such tree level corrections in the
Supernova context have been partially studied in
\cite{Amanik:2004vm,ro91}. Since this work focuses on $V_{\mu
\tau}$, we only concentrate on graphs involving $\mu$ and $\tau$
neutrinos. Moreover, we only consider neutrinos which do not
change of flavour after the R-parity breaking interaction,
therefore we are not concerned by
off-diagonal term in the matter Hamiltonian. \\

\begin{figure}[h]
\vspace{.6cm}
\centerline{\includegraphics[scale=0.3,angle=0]{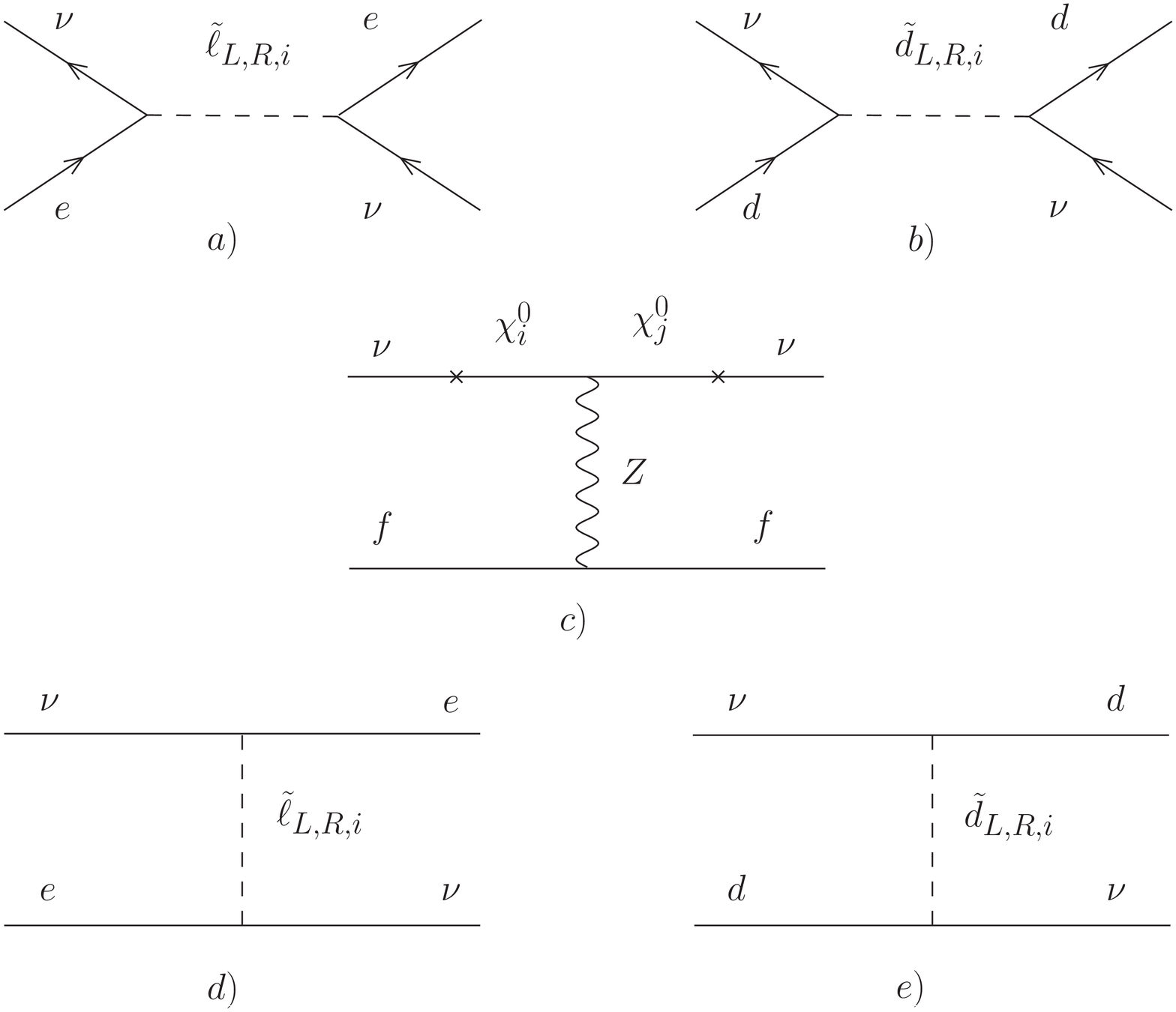}\hspace{.2cm}}
\caption{R-parity broken Tree-level corrections.} \label{fig:RP6}
\end{figure}

Here, we give a specific contribution to antineutrino scattering
(Fig.\ref{fig:RP6}a) with left sleptons $\tilde{\ell}_{L,i}$:
\begin{equation}
\Delta\rho_a(\tilde{\ell}_{L,i}) =
-\sum_{i=1}^{3}\frac{(\lambda_{3i1}^2 -
\lambda_{2i1}^2)}{g^2}\frac{M_W^2}{m_{\tilde{\ell}_{L,i}}^2}
\end{equation}

With left down squarks $\tilde{d}_{L,i}$ (Fig.\ref{fig:RP6}b):
\begin{equation}
\Delta\rho_b(\tilde{d}_{L,i}) = \Delta\rho_a(\tilde{\ell}_{L,i}) \
(\lambda_{ki1}\rightarrow \lambda_{ki1}',
\tilde{\ell}_{L,i}\rightarrow \tilde{d}_{L,i})
\end{equation}

The right sleptons $\tilde{\ell}_{R,i}$ and the right down squarks
$\tilde{d}_{R,i}$ do not contribute.

The following contributions induce corrections to both neutrino
and antineutrino scattering:

for diagram (Fig.\ref{fig:RP6}c):
\begin{eqnarray}\label{treelevelchi0}
\Delta\rho_c(\chi^0)=-\sum_{j,k=1}^5\left(N_{(j+3)8}N_{(i+3)8} -
N_{(j+3)7}N_{(i+3)7} \right)\left(N_{(j+3)4}N_{(i+3)4} -
N_{(j+3)3}N_{(i+3)3} \right)
\end{eqnarray}

for diagram (Fig.\ref{fig:RP6}d):
\begin{equation}
\Delta\rho_d(\tilde{\ell}_{L,i}) =
\Delta\rho_a(\tilde{\ell}_{L,i})
\end{equation}

for diagram (Fig.\ref{fig:RP6}e):
\begin{equation}
\Delta\rho_e(\tilde{d}_{L,i}) = \Delta\rho_b(\tilde{d}_{L,i})
\end{equation}

For (Fig.\ref{fig:RP6}d) and (Fig.\ref{fig:RP6}e), the right
sleptons $\tilde{\ell}_{R,i}$ and the right down squarks
$\tilde{d}_{R,i}$ do not contribute.

\subsection{Self-energy corrections}

We willingly leave certain factors not simplified to be able to
compare with other corrections more rapidly. All the contributions
will be equal for neutrino and antineutrino scattering.

\begin{figure}[h]
\vspace{.6cm}
\centerline{\includegraphics[scale=0.3,angle=0]{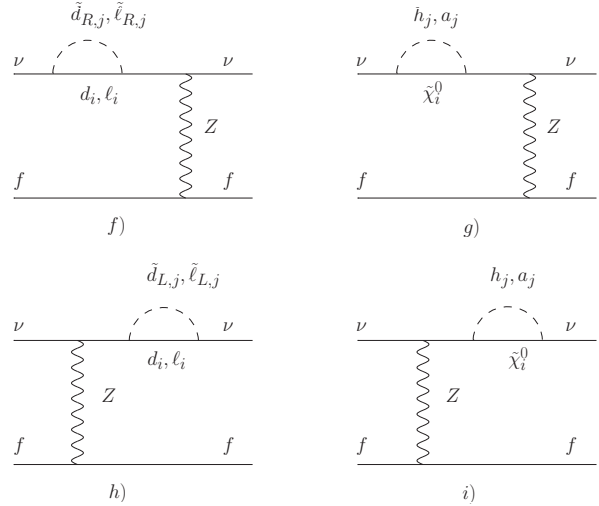}\hspace{.2cm}}
\caption{R-parity broken Self-energy corrections. }
\label{fig:RP2}
\end{figure}

For Fig.(\ref{fig:RP2}f):

\begin{eqnarray}\label{Self3}
 \Delta\rho^{\tilde{\ell}_R}(\Sigma)=-{\alpha_W\over 8\pi}
\sum_{i,j=1}^3
\frac{(\lambda_{3ij}^2-\lambda_{2ij}^2)}{g^2}\left\{
G_0(X_{\ell_i\tilde{\ell}_{R,j}},1)+{\rm ln}
{m_{\tilde{\ell}_{R,j}}^2\over \mu^2} \right\},
\end{eqnarray}

$$\Delta\rho^{\tilde{d}_R}(\Sigma) =
 \Delta\rho^{\tilde{\ell}_R}(\Sigma)\
(\lambda_{kij}\rightarrow \lambda_{kij}',
\tilde{\ell}_{R,j}\rightarrow \tilde{d}_{R,j})$$\label{Self4}

For Fig.(\ref{fig:RP2}h):

\begin{eqnarray}\label{Self5}
 \Delta\rho^{\tilde{\ell}_L}(\Sigma) =
 \Delta\rho^{\tilde{\ell}_R}(\Sigma) \
(\lambda_{kij}\rightarrow \lambda_{kji},
\tilde{\ell}_{R,j}\rightarrow \tilde{\ell}_{L,j}),
\end{eqnarray}

$$\Delta\rho^{\tilde{d}_L}(\Sigma) =
 \Delta\rho^{\tilde{\ell}_R}(\Sigma) \
(\lambda_{kij}\rightarrow \lambda_{kji}',
\tilde{\ell}_{R,j}\rightarrow \tilde{d}_{L,j})$$\label{Self6}

In the R-parity breaking scenario, there are NMSSM specific
self-energies due to the term
$\lambda_{i}\hat{S}\hat{H}_{u}\hat{L}_{i}$ in the superpotential.
This term induce one new coupling between a neutrino, one Higgs
(scalar or pseudoscalar) and a neutralino. We present here the
contributions for scalar Higgses and for pseudoscalar Higgses.

For both Fig.(\ref{fig:RP2}g) and (\ref{fig:RP2}i), the
contribution from scalar higgses:
\begin{eqnarray}\label{Self7}
 \Delta\rho^{h}(\Sigma)=-{\alpha_W\over 8\pi}
 \frac{(\lambda_3^2-\lambda_2^2)}{g^2}
\sum_{i=1}^5\sum_{j=1}^3
(N_{(i+3)5}^2S_{j1}^2+N_{(i+3)3}^2S_{j3}^2)\left\{ G_0(X_{\chi^0_i
h_j},1)+{\rm ln} {m_{h_j}^2\over \mu^2} \right\}
\end{eqnarray}

from pseudoscalar higgses:
\begin{eqnarray}\label{Self8}
 \Delta\rho^{a}(\Sigma)=-{\alpha_W\over 8\pi}
 \frac{(\lambda_3^2-\lambda_2^2)}{g^2}
\sum_{i=1}^5\sum_{j=1}^2
(N_{(i+3)5}^2P_{j1}^2+N_{(i+3)3}^2P_{j3}^2)\left\{ G_0(X_{\chi^0_i
a_j},1)+{\rm ln} {m_{a_j}^2\over \mu^2} \right\}
\end{eqnarray}

\subsection{penguin type diagrams}

\begin{figure}[h]
\vspace{.6cm}
\centerline{\includegraphics[scale=0.3,angle=0]{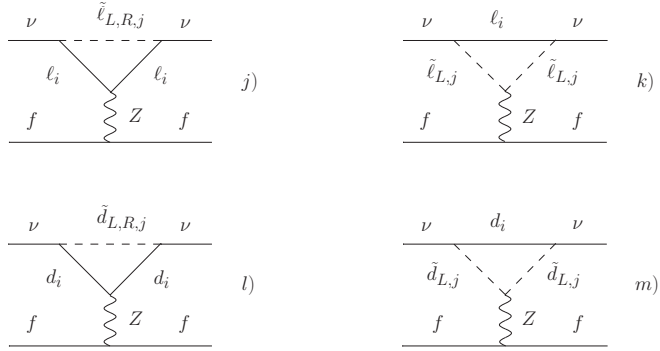}\hspace{.2cm}}
\caption{R-parity broken penguin type diagrams.} \label{fig:RP5}
\end{figure}

We consider here the penguin diagrams. All the contributions will
be equal for neutrino and antineutrino scattering.

For the diagram of Fig.(\ref{fig:RP5}j) and Fig.(\ref{fig:RP5}l):
\begin{equation}
\Delta\rho_P^{\tilde{\ell}_L}(\ell) = -\frac{\alpha_W}{16\pi}
\sum_{i,j=1}^3
\frac{(\lambda^2_{3ji}-\lambda^2_{2ji})}{g^2}\left[G_0(X_{\tilde{\ell}_{L,j},\ell_i},1)+\ln(\frac{m^2_{{\ell}_i}}{\mu^2})\right],
\end{equation}

$$\Delta\rho_P^{\tilde{d}_L}(d) =
\Delta\rho_P^{\tilde{\ell}_L}(\ell) \ (\lambda_{kji}\rightarrow
\lambda_{kji}', \tilde{\ell}_{L,j}\rightarrow \tilde{d}_{L,j},
\ell_i \rightarrow d_i)$$

For the diagram of Fig.(\ref{fig:RP5}k) and Fig.(\ref{fig:RP5}m):
\begin{equation}
\Delta\rho_P^{\ell}(\tilde{\ell}_L) =
-\frac{\alpha_W}{4\pi}c^{\ell}_L \sum_{i,j=1}^3
\frac{(\lambda^2_{3ji}-\lambda^2_{2ji})}{g^2}\left[G_0(X_{\ell_i,\tilde{\ell}_{L,j}},1)+\ln(\frac{m^2_{\tilde{\ell}_{L,j}}}{\mu^2})
\right]
\end{equation}

$$\Delta\rho_P^{d}(\tilde{d}_L) =
\Delta\rho_P^{\ell}(\tilde{\ell}_L) \ (\lambda_{kji}\rightarrow
\lambda_{kji}', \tilde{\ell}_{L,j}\rightarrow \tilde{d}_{L,j},
\ell_i \rightarrow d_i)$$

For all the previous penguin diagrams, the right sleptons
$\tilde{\ell}_{R,j}$ and the right squarks $\tilde{d}_{R,j}$ do
not contribute.

We have below NMSSM specific penguins also due to the term
$\lambda_{i}\hat{S}\hat{H}_{u}\hat{L}_{i}$ in the superpotential.

\begin{figure}[h]
\vspace{.6cm}
\centerline{\includegraphics[scale=0.3,angle=0]{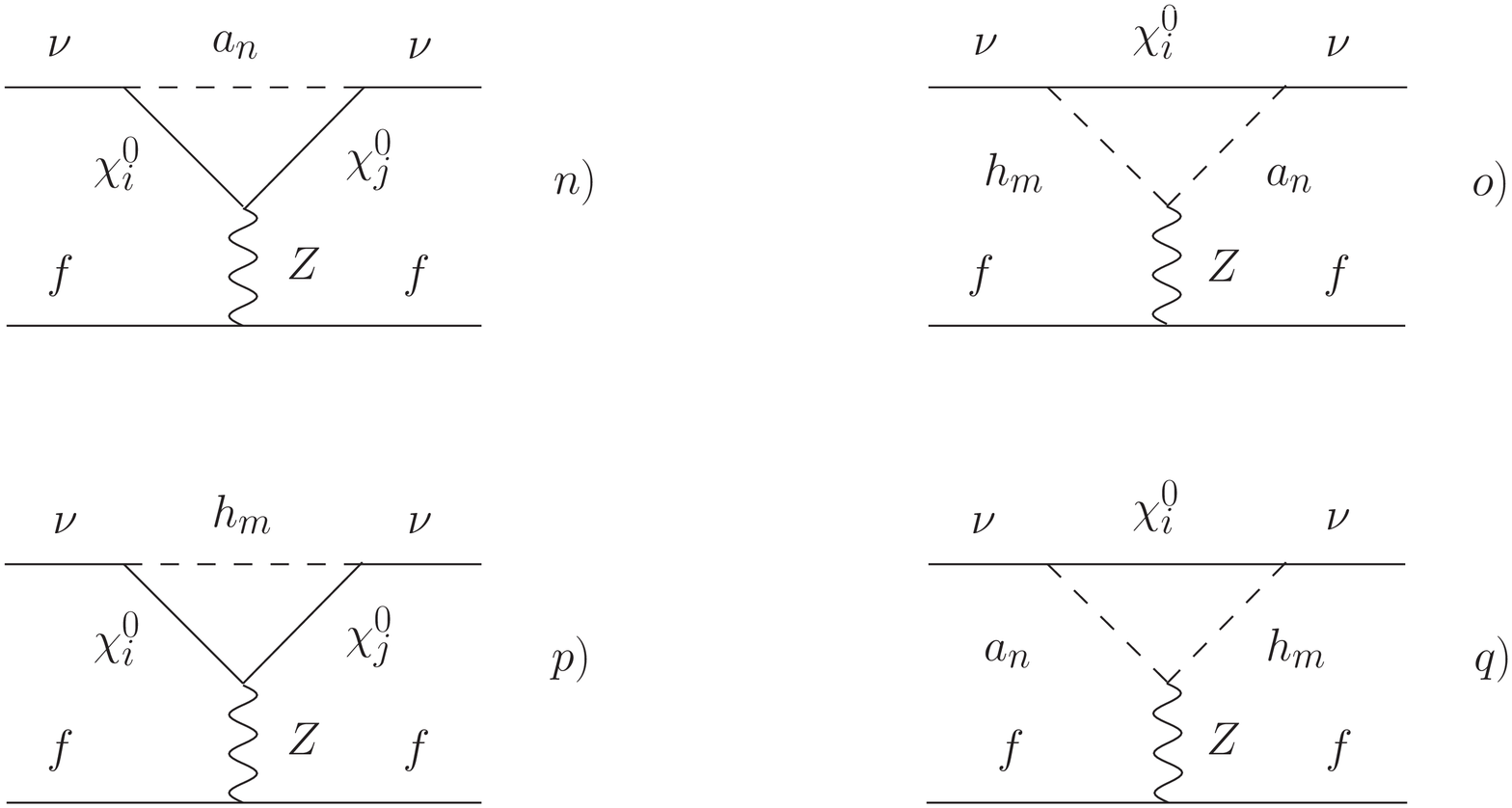}\hspace{.2cm}}
\caption{R-parity broken NMSSM specific penguin type diagrams.}
\label{fig:RP5}
\end{figure}

For the diagram of Fig.(\ref{fig:RP5}p):
\begin{equation}\label{NMSSMspecifichm}
\begin{array}{rcl}
\Delta\rho^h_p(\chi^0)&=&-{\alpha_W\over 8\pi
}\frac{\lambda_3^2-\lambda_2^2}{g^2} \sum_{i,j=1}^5\sum_{m=1}^3
N_{(i+3)5}N_{(j+3)5} S_{m1}^{2}\left(N_{(j+3)4}N_{(i+3)4}
- N_{(j+3)3}N_{(i+3)3} \right)  \\
&\ & \times \left[2H_0(X_{\chi^0_i h_{m}},X_{\chi^0_j h_{m}}) +
\left\{G_0(X_{\chi^0_i h_{m}},X_{\chi^0_j h_{m}})+{\rm
ln}{m_{h_{m}}^2\over\mu^2}\right\} \right] ,
\end{array}
\end{equation}

For the diagram of Fig.(\ref{fig:RP5}n):
\begin{equation}\label{NMSSMspecifican}
\begin{array}{rcl}
\Delta\rho^a_p(\chi^0)&=&-{\alpha_W\over 8\pi
}\frac{\lambda_3^2-\lambda_2^2}{g^2} \sum_{i,j=1}^5\sum_{n=1}^2
N_{(i+3)5}N_{(j+3)5} P_{n1}^{2}\left(N_{(j+3)4}N_{(i+3)4}
- N_{(j+3)3}N_{(i+3)3} \right)  \\
&\ & \times \left[2H_0(X_{\chi^0_i a_{n}},X_{\chi^0_j a_{n}}) +
\left\{G_0(X_{\chi^0_i a_{n}},X_{\chi^0_j a_{n}})+{\rm
ln}{m_{a_{n}}^2\over\mu^2}\right\} \right] ,
\end{array}
\end{equation}

We have two identical contributions from the diagram of
Fig.(\ref{fig:RP5}o) and Fig.(\ref{fig:RP5}q) and we obtain:
\begin{eqnarray}\label{NMSSMspecificchi0}
\Delta\rho^{\chi^0}_p=-{\alpha_W\over 4\pi
}\frac{\lambda_3^2-\lambda_2^2}{g^2}
\sum_{i=1}^5\sum_{m=1}^3\sum_{n=1}^2 N_{(i+3)5}^{2}
S_{m1}P_{n1}A^{mn}_M
\left\{G_0(X_{h_{m}\chi^0_i},X_{a_{n}\chi^0_i})+{\rm
ln}{m_{\chi^0_i}^2\over\mu^2}\right\}
\end{eqnarray}
where $A^{mn}_{\it{M}} = S_{m1}P_{n1} - S_{m2}P_{n2}$ following
the notation of \cite{nmssmtools1}.

\subsection{Box diagrams}

\subsubsection{R-parity broken box diagrams}

\begin{figure}
\vspace{.6cm}
\centerline{\includegraphics[scale=0.3,angle=0]{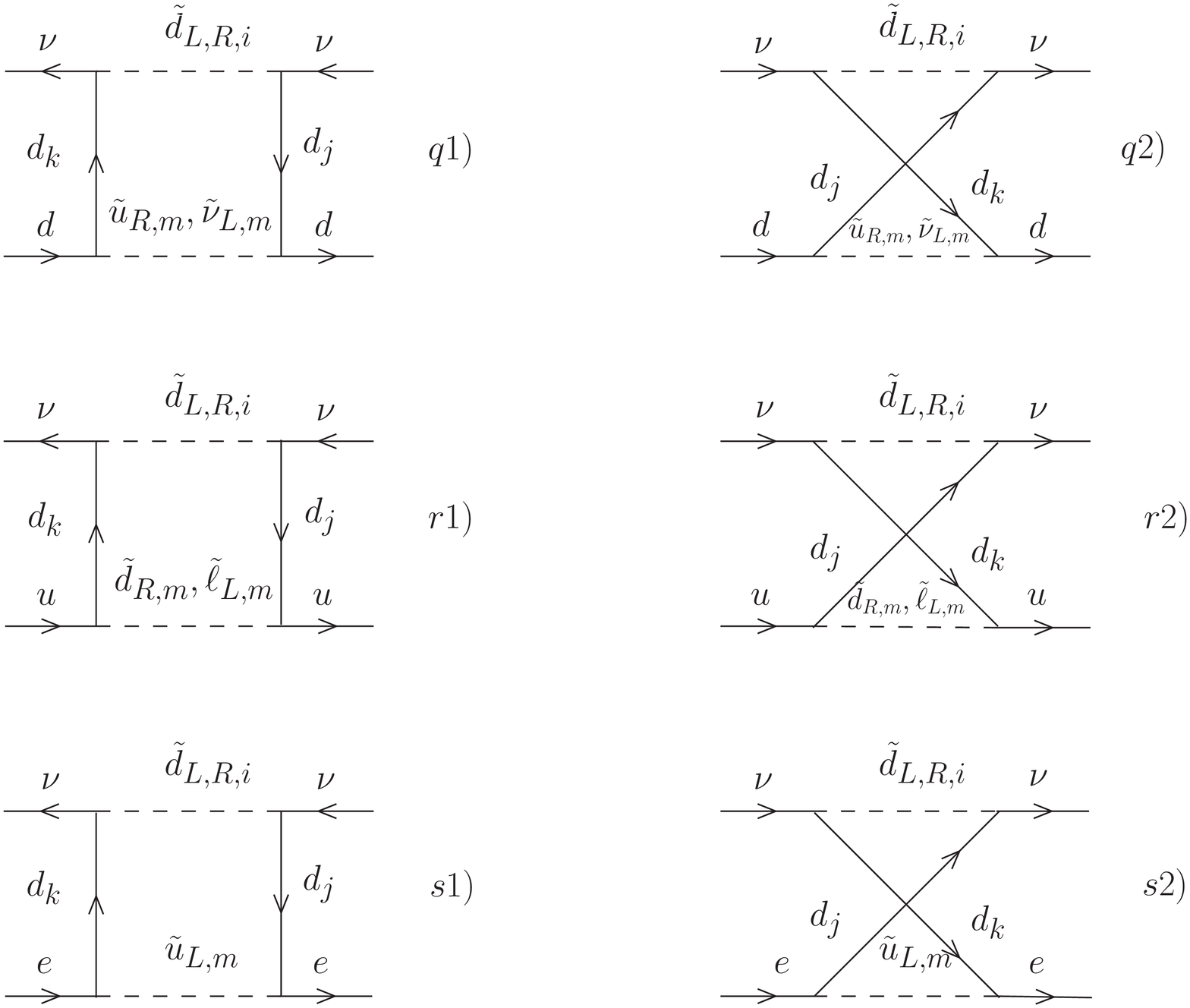}\hspace{.2cm}
\includegraphics[scale=0.3,angle=0]{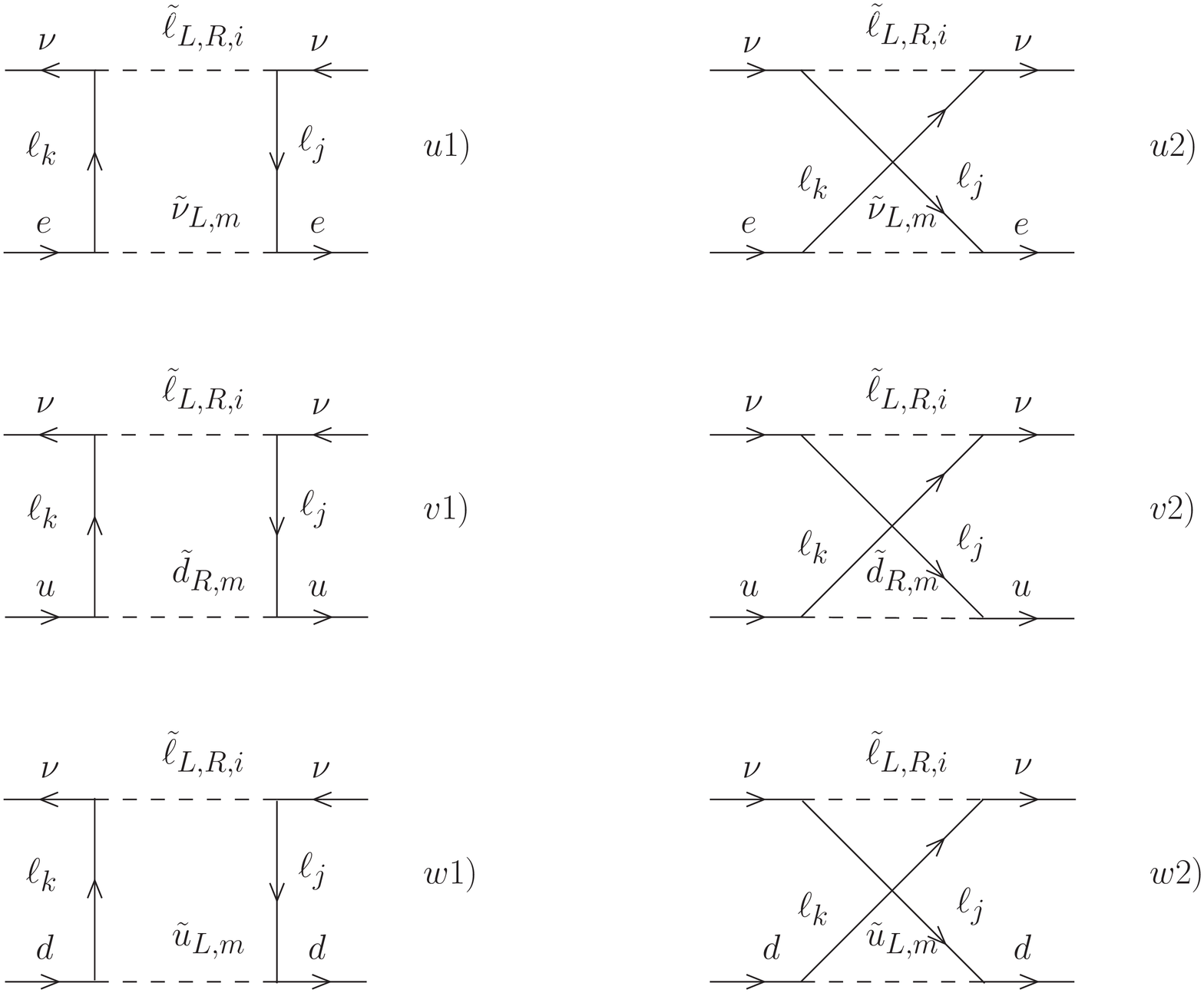}}
\caption{R-parity broken box diagrams.} \label{fig:RP4}
\end{figure}

We present in this section the box diagrams corrections. Each box
diagram correction involves Fierz transformations to obtain a
similar form as the SM tree level interaction. The box diagrams
with crossed fermions lines (Fig.\ref{fig:RP4}q2,
Fig.\ref{fig:RP4}r2, Fig.\ref{fig:RP4}s2, Fig.\ref{fig:RP4}u2,
Fig.\ref{fig:RP4}v2, Fig.\ref{fig:RP4}w2) will only contribute to
neurinos, meanwhile the uncrossed fermion lines
(Fig.\ref{fig:RP4}q1, Fig.\ref{fig:RP4}r1, Fig.\ref{fig:RP4}s1,
Fig.\ref{fig:RP4}u1, Fig.\ref{fig:RP4}v1, Fig.\ref{fig:RP4}w1)
will contribute to antineutrinos. However, the contributions to
antineutrino scattering will be exactly the same as the
contributions to neutrino scattering. We show below the
contributions to neutrino scattering.

For the diagram of Fig.(\ref{fig:RP4}v2) with
$\tilde{\ell}_{L,i}-\tilde{d}_{R,m}$:
\begin{equation}
\Delta\rho_{box}^{v2}(\tilde{\ell}_{L}-\tilde{d}_{R})=
-\frac{\alpha_W}{8\pi} \sum_{i,j,k,m=1}^{3}
\frac{M^2_W}{m^2_{\tilde{d}_{R,m}}}
\frac{(\lambda_{3ij}\lambda_{3ik}-\lambda_{2ij}\lambda_{2ik})\lambda'_{j1m}\lambda'_{k1m}}{g^4}
G'(X_{\ell_j,\tilde{d}_{R,m}},X_{\ell_k,\tilde{d}_{R,m}},X_{\tilde{\ell}_{L,i},\tilde{d}_{R,m}}).
\end{equation}

For the diagram of Fig.(\ref{fig:RP4}r2) with
$\tilde{d}_{L,i}-\tilde{\ell}_{L,m}$ and the diagram of
Fig.(\ref{fig:RP4}s2) with $\tilde{d}_{L,i}-\tilde{u}_{L,m}$:
\begin{equation}
\Delta\rho_{box}^{r2}(\tilde{d}_{L}-\tilde{\ell}_{L}) =
\Delta\rho_{box}^{v2}(\tilde{\ell}_{L}-\tilde{d}_{R}) \
(\lambda\rightarrow \lambda',
\lambda'_{j1m}\lambda'_{k1m}\rightarrow
\lambda'_{m1j}\lambda'_{m1k}, l_{j,k}\rightarrow d_{j,k},
\tilde{d}_{R,m}\rightarrow \tilde{\ell}_{L,m}),
\end{equation}

$$\Delta\rho_{box}^{s2}(\tilde{d}_{L}-\tilde{u}_{L}) = \Delta\rho_{box}^{r2}(\tilde{d}_{L}-\tilde{\ell}_{L})
\ (\lambda'_{m1j}\lambda'_{m1k}\rightarrow
\lambda'_{1mj}\lambda'_{1mk}, \tilde{\ell}_{L,m}\rightarrow
\tilde{u}_{L,m}).$$

For the diagram of Fig.(\ref{fig:RP4}r2) with
$\tilde{d}_{R,i}-\tilde{\ell}_{L,m}$ and the diagram of
Fig.(\ref{fig:RP4}s2) with $\tilde{d}_{R,i}-\tilde{u}_{L,m}$:
\begin{equation}
\Delta\rho_{box}^{r2}(\tilde{d}_{R}-\tilde{\ell}_{L})=
\frac{\alpha_W}{4\pi} \sum_{i,j,k,m=1}^{3}
\frac{M^2_W}{m^2_{\tilde\ell_{L,m}}}
\frac{(\lambda'_{3ji}\lambda'_{3ki}-\lambda'_{2ji}\lambda'_{2ki})\lambda'_{m1j}\lambda'_{m1k}}{g^4}
H'(X_{d_j,\tilde\ell_{L,m}},X_{d_k,\tilde\ell_{L,m}},X_{\tilde{d}_{R,i},\tilde\ell_{L,m}}),
\end{equation}

$$\Delta\rho_{box}^{s2}(\tilde{d}_{R}-\tilde{u}_{L}) = \Delta\rho_{box}^{r2}(\tilde{d}_{R}-\tilde{\ell}_{L})
\ (\lambda'_{m1j}\lambda'_{m1k}\rightarrow
\lambda'_{1mj}\lambda'_{1mk}, \tilde{\ell}_{L,m}\rightarrow
\tilde{u}_{L,m}).$$

For the diagram of Fig.(\ref{fig:RP4}q2) with
$\tilde{d}_{R,i}-\tilde{\nu}_{L,m}$ and for the diagram of
Fig.(\ref{fig:RP4}u2) with $\tilde{\ell}_{R,i}-\tilde{\nu}_{L,m}$:
\begin{equation}
\Delta\rho_{box}^{q2}(\tilde{d}_{R}-\tilde{\nu}_{L})=
\frac{\alpha_W}{4\pi} \sum_{i,j,k,m=1}^{3}
\frac{M^2_W}{m^2_{\tilde{\nu}_{L,m}}}
\frac{(\lambda'_{3ij}\lambda'_{3ik}-\lambda'_{2ij}\lambda'_{2ik})\lambda'_{mj1}\lambda'_{mk1}}{g^4}
H'(X_{d_j,\tilde{\nu}_{L,m}},X_{d_k,\tilde{\nu}_{L,m}},X_{\tilde{d}_{R,i},\tilde{\nu}_{L,m}}),
\end{equation}

$$\Delta\rho_{box}^{u2}(\tilde{\ell}_{R}-\tilde{\nu}_{L}) =
\Delta\rho_{box}^{q2}(\tilde{d}_{R}-\tilde{\nu}_{L}) \
(\lambda'_{mj1}\lambda'_{mk1}\rightarrow
\lambda'_{m1j}\lambda'_{m1k} {\rm then} \lambda' \rightarrow
\lambda, d_{j,k} \rightarrow \ell_{j,k}, \tilde{d}_{R,i}
\rightarrow \tilde{\ell}_{R,i}).$$

For the diagram of Fig.(\ref{fig:RP4}v2) with
$\tilde{\ell}_{R,i}-\tilde{d}_{R,m}$ and for the diagram of
Fig.(\ref{fig:RP4}w2) with $\tilde{\ell}_{L,i}-\tilde{u}_{L,m}$:
\begin{equation}
\Delta\rho_{box}^{v2}(\tilde{\ell}_{R}-\tilde{d}_{R})=
\frac{\alpha_W}{4\pi} \sum_{i,j,k,m=1}^{3}
\frac{M^2_W}{m^2_{\tilde{d}_{R,m}}}
\frac{(\lambda_{3ji}\lambda_{3ki}-\lambda_{2ji}\lambda_{2ki})\lambda'_{j1m}\lambda'_{k1m}}{g^4}
H'(X_{\ell_j,\tilde{d}_{R,m}},X_{\ell_k,\tilde{d}_{R,m}},X_{\tilde{\ell}_{R,i},\tilde{d}_{R,m}}),
\end{equation}

$$\Delta\rho_{box}^{w2}(\tilde{\ell}_{L}-\tilde{u}_{L}) = \Delta\rho_{box}^{v2}(\tilde{\ell}_{R}-\tilde{d}_{R}) \
(\lambda'_{j1m}\lambda'_{k1m}\rightarrow
\lambda'_{jm1}\lambda'_{km1}, \tilde{d}_{R,m} \rightarrow
\tilde{u}_{L,m}, \tilde{\ell}_{R,i} \rightarrow
\tilde{\ell}_{L,i}).$$

For the diagram of Fig.(\ref{fig:RP4}r2) with
$\tilde{d}_{L,i}-\tilde{d}_{R,m}$:
\begin{equation}
\Delta\rho_{box,\alpha\sigma}^{r2}(\tilde{d}_{L}-\tilde{d}_{R})=
\frac{\alpha_W}{4\pi} \sum_{i,j,k,m=1}^{3}
\sum_{\beta,\delta,\gamma =1}^{3}
\frac{M^2_W}{m^2_{\tilde{d}^\gamma_{R,m}}}
\frac{(\lambda'_{3ij}\lambda'_{3ik}-\lambda'_{2ij}\lambda'_{2ik})\lambda''_{1mj}\lambda''_{1mk}}{g^4}
\epsilon_{\sigma\gamma\delta}\epsilon_{\alpha\gamma\beta}
H'(X_{d^\delta_j,\tilde{d}^\gamma_{R,m}},X_{d^\beta_k,\tilde{d}^\gamma_{R,m}},X_{\tilde{d}_{L,i},\tilde{d}^\gamma_{R,m}}).
\end{equation}

For the diagram of Fig.(\ref{fig:RP4}q2) with
$\tilde{d}_{R,i}-\tilde{u}_{R,m}$:
\begin{equation}
\Delta\rho_{box,\beta\sigma}^{q2}(\tilde{d}_{R}-\tilde{u}_{R})=
\frac{\alpha_W}{16\pi} \sum_{i,j,k,m=1}^{3}
\sum_{\alpha,\delta,\gamma =1}^{3}
\frac{M^2_W}{m^2_{\tilde{u}^\alpha_{R,m}}}
\frac{(\lambda'_{3ji}\lambda'_{3ki}-\lambda'_{2ji}\lambda'_{2ki})\lambda''_{mj1}\lambda''_{mk1}}{g^4}
\epsilon_{\alpha\delta\sigma}\epsilon_{\alpha\gamma\beta}
H'(X_{d_j^\gamma,\tilde{u}^\alpha_{R,m}},X_{d_k^\delta,\tilde{u}^\alpha_{R,m}},X_{\tilde{d}_{R,i},\tilde{u}^\alpha_{R,m}}).
\end{equation}

The indices of $\epsilon$ denote $SU(3)$ color indices where
$\epsilon$ is antisymmetric, $\epsilon_{\alpha\beta\gamma} =
\epsilon_{\beta\gamma\alpha} = \epsilon_{\gamma\alpha\beta}$. For
example, for the diagram of Fig.(\ref{fig:RP4}q2) with
$\tilde{d}_{L,i}-\tilde{u}_{R,m}$, the down quark on the left
carries the index $\beta$ and the down quark on the right carries
the index $\sigma$:
\begin{eqnarray}
\Delta\rho_{box,\beta\sigma}^{q2}(\tilde{d}_{L}-\tilde{u}_{R}) &=&
\frac{\alpha_W}{16\pi} \sum_{i,j,k,m=1}^{3}
\sum_{\alpha,\delta,\gamma =1}^{3}
\frac{M^2_W}{m^2_{\tilde{u}^\alpha_{R,m}}}
\frac{(\lambda'_{3ij}\lambda'_{3ik}-\lambda'_{2ij}\lambda'_{2ik})
\lambda''_{m1j}\lambda''_{m1k}}{g^4}\epsilon_{\alpha\sigma\delta}\epsilon_{\alpha\beta\gamma} \\
&&
\left[H'(X_{d_j^\delta,\tilde{u}^\alpha_{R,m}},X_{d_k^\gamma,\tilde{u}^\alpha_{R,m}},X_{\tilde{d}_{L,i},\tilde{u}^\alpha_{R,m}})
-\frac{G'(X_{d_j^\delta,\tilde{u}^\alpha_{R,m}},X_{d_k^\gamma,\tilde{u}^\alpha_{R,m}},X_{\tilde{d}_{L,i},\tilde{u}^\alpha_{R,m}})}
{2} \right]. \nonumber
\end{eqnarray}

The diagram of Fig.(\ref{fig:RP4}q2) with
$\tilde{d}_{L,i}-\tilde{\nu}_{L,m}$ can be deducted from the
previous one by removing the $\epsilon$ factors and by taking a
factor 4 and $\lambda'' \rightarrow \lambda'$, $\tilde{u}_R
\rightarrow \tilde{\nu}_L$:
\begin{eqnarray}
\Delta\rho_{box}^{q2}(\tilde{d}_{L}-\tilde{\nu}_{L}) &=&
\frac{\alpha_W}{4\pi} \sum_{i,j,k,m=1}^{3}
\frac{M^2_W}{m^2_{\tilde{\nu}_{L,m}}}
\frac{(\lambda'_{3ij}\lambda'_{3ik}-\lambda'_{2ij}\lambda'_{2ik})
\lambda'_{mj1}\lambda'_{mk1}}{g^4} \\
&&
\left[H'(X_{d_j,\tilde{\nu}_{L,m}},X_{d_k,\tilde{\nu}_{L,m}},X_{\tilde{d}_{L,i},\tilde{\nu}_{L,m}})
-\frac{G'(X_{d_j,\tilde{\nu}_{L,m}},X_{d_k,\tilde{\nu}_{L,m}},X_{\tilde{d}_{L,i},\tilde{\nu}_{L,m}})}
{2}\right]. \nonumber
\end{eqnarray}

Finally, for the diagram of Fig.(\ref{fig:RP4}u2) with
$\tilde{\ell}_{L,i}-\tilde{\nu}_{L,m}$:
\begin{eqnarray}
\Delta\rho_{box}^{u2}(\tilde{\ell}_{L}-\tilde{\nu}_{L}) =
\Delta\rho_{box}^{q2}(\tilde{d}_{L}-\tilde{\nu}_{L}) \ (\lambda'
\rightarrow \lambda, d_{j,k} \rightarrow \ell_{j,k},
\tilde{d}_{L,i} \rightarrow \tilde{\ell}_{L,i}).
\end{eqnarray}

The rest of diagrams do not contribute because of a different form
as the correct tree level SM form given in
Eq.(\ref{SMscatteringamplitude}).

\subsubsection{Cancelling diagrams}

\begin{figure}[h]
\vspace{.6cm}
\centerline{\includegraphics[scale=0.3,angle=0]{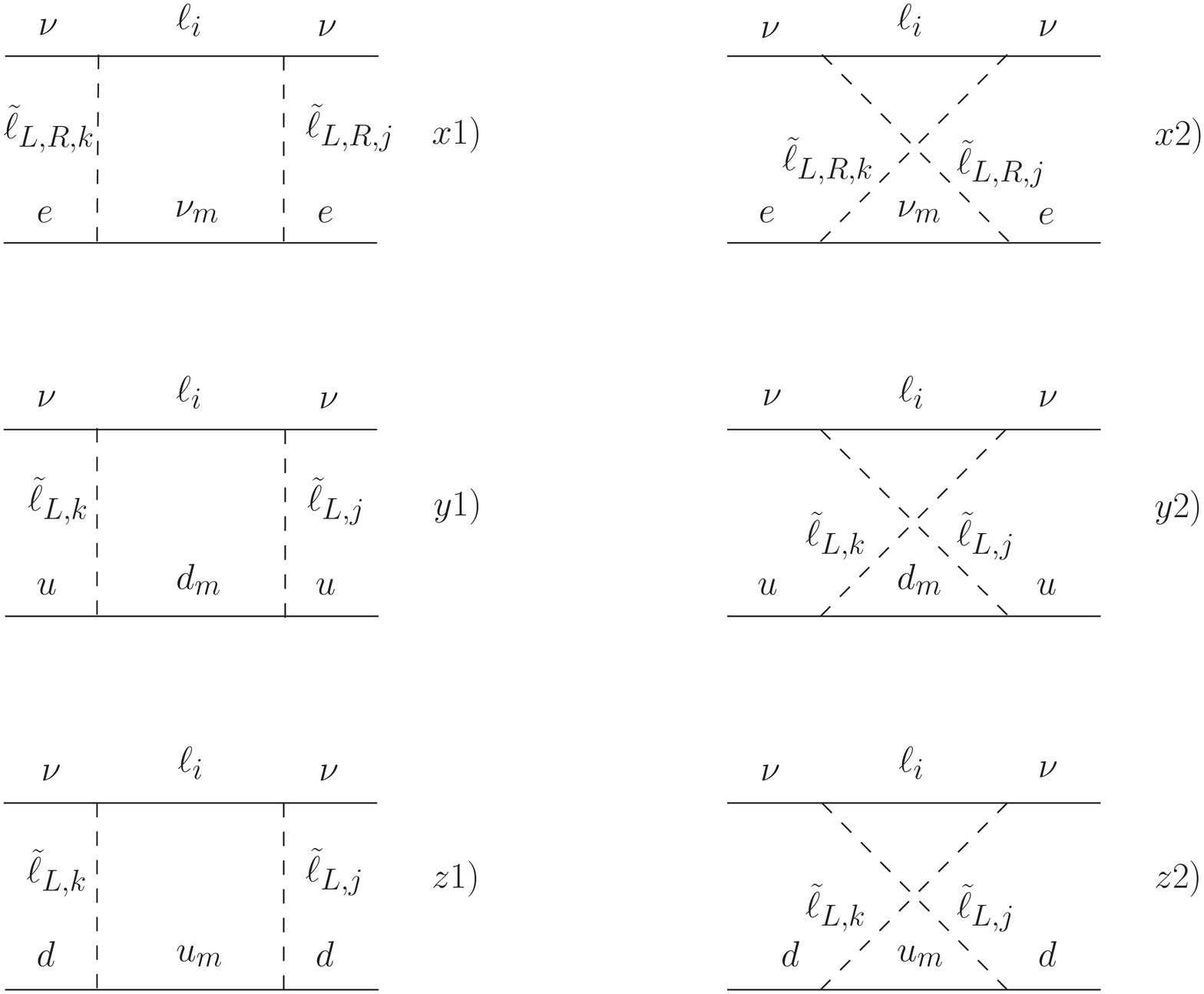}\hspace{.2cm}
\includegraphics[scale=0.3,angle=0]{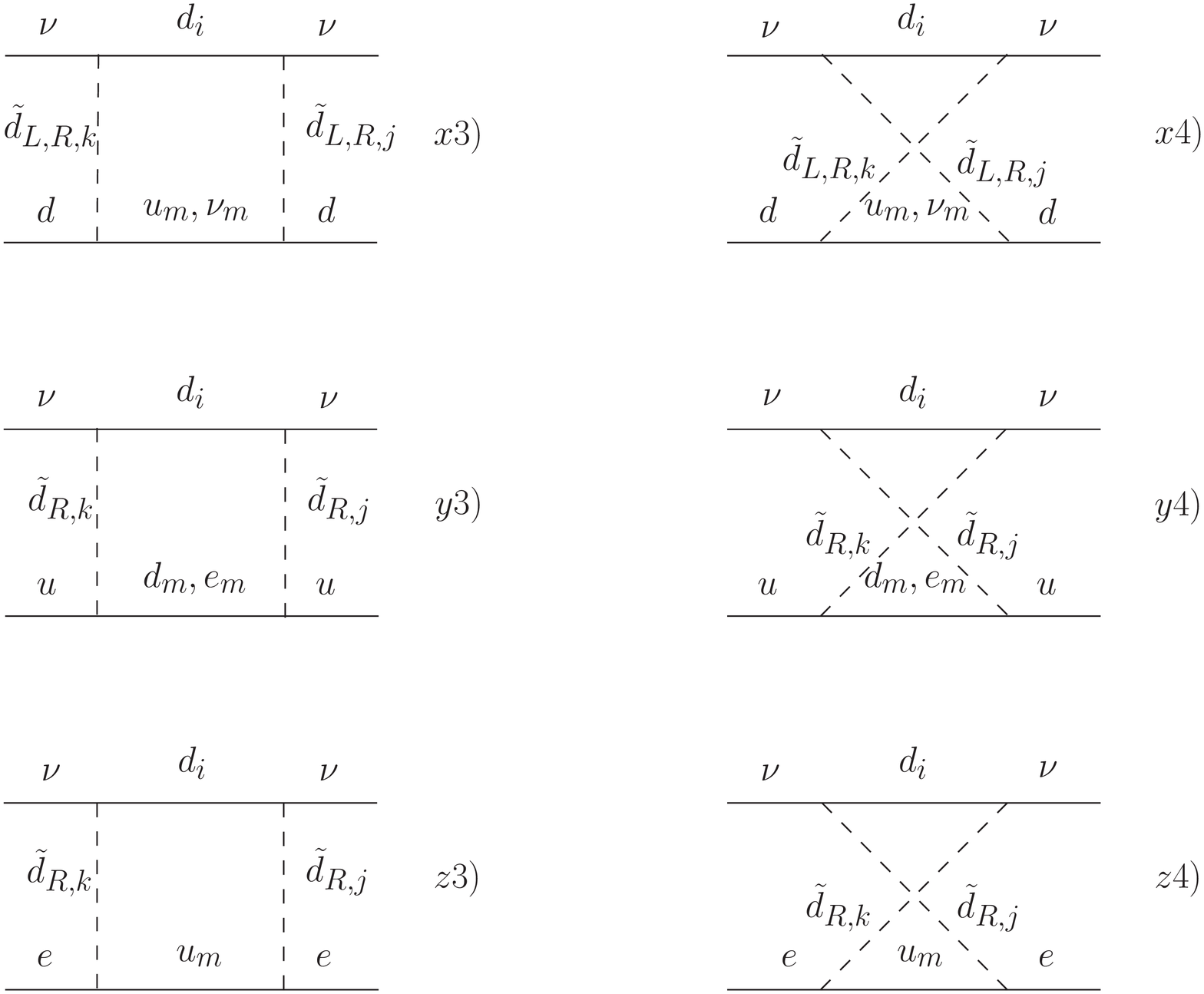}}
\caption{R-parity broken radiative corrections which cancel out.
The Feynman graphs with uncrossed scalar lines on the left cancel
with the Feynman graphs with crossed scalar lines on the right.}
\label{fig:RP1e3}
\end{figure}
It is interesting to notice that amongst the possible radiative
corrections induced by the R-parity breaking interactions, some of
them cancel. Contrary to the previous box diagrams, all box
diagrams in Fig.(\ref{fig:RP1e3}) should contribute equally to
neutrinos and anti-neutrinos because the fermionic lines do not
link the neutrinos to the fermion present in the external leg. The
particles exchanged between the two fermionic lines here are all
scalar.

To understand the consequences, we write the tensorial part of the
scattering amplitude of a ladder scalar line box diagram
(Fig.\ref{fig:RP1e3}x1) and the corresponding crossed scalar line
box diagram (Fig.\ref{fig:RP1e3}x2):

For the left diagram we have: \ba {\cal{M}}_{1} & = &
\bar{u}(k_1)\left[\int \frac{dq^4}{(2\pi)^4}i
\lambda_{3ki}\frac{1+\gamma_5}{2}\frac{i(\slashed{q}+m_{\ell_i})}{q^2-m^2_{\ell_i}}
i \lambda_{3ji}\frac{1-\gamma_5}{2}u(p_1)\frac{i}{(p_1-q)^2-m^2_{\tilde{\ell}_{L,j}}}\frac{i}{(k_1-q)^2-m^2_{\tilde{\ell}_{L,k}}}\right. \nonumber \\
&& \left.\bar{u}(k_2)i
\lambda_{mk1}\frac{1-\gamma_5}{2}\frac{i(\slashed{p_2}+\slashed{p_1}-\slashed{q}+m_{\nu_m})}{q^2-m^2_{\nu_m}}
i\lambda_{mj1}\frac{1+\gamma_5}{2}\right]u(p_2) \ea

while for the right diagram we have: \ba {\cal{M}}_{2} & = &
\bar{u}(k_1)\left[\int \frac{dq^4}{(2\pi)^4}i
\lambda_{3ji}\frac{1+\gamma_5}{2}\frac{i(\slashed{q}+m_{\ell_i})}{q^2-m^2_{\ell_i}}
i \lambda_{3ki}\frac{1-\gamma_5}{2}u(p_1)\frac{i}{(p_1-q)^2-m^2_{\tilde{\ell}_{L,k}}}\frac{i}{(k_1-q)^2-m^2_{\tilde{\ell}_{L,j}}}\right. \nonumber \\
&& \left.\bar{u}(k_2)i
\lambda_{mk1}\frac{1-\gamma_5}{2}\frac{i(\slashed{p_2}-\slashed{k_1}+\slashed{q}+m_{\nu_m})}{q^2-m^2_{\nu_m}}
i\lambda_{mj1}\frac{1+\gamma_5}{2}\right]u(p_2) \ea

Therefore , in the approximation of zero external legs ($p_1$,
$p_2$, $k_1$, $k_2$ = 0), we have ${\cal{M}}_{1}=-{\cal{M}}_{2}$.
From a physical point of view, it is actually quite natural that
such corrections cancel out when considering vanishing external
legs, the Feynman box diagrams are anti-symmetric in this case
since scalar particles are exchanged and the sign of the $\nu_m$
neutrino impulsion is opposite.

\section{Implications on the supernova neutrino fluxes}

In the supernova environment the density is sufficiently high that
neutrinos encounter not only the so called high resonance
associated with $\theta_{13}$ and the low resonance associated
with $\theta_{12}$ but also the $\mu- \tau$ resonance whose
precise conditions depend on the hierarchy \cite{Kneller:2009vd}.
The importance of such correction term has been investigated in
\cite{EstebanPretel:2007yq,Duan:2008za} showing that such
resonance could influence the electron neutrino flux. This
radiative correction term may also influence the electron (anti-)
neutrino in an indirect way. Indeed, $V_{\mu \tau}$ breaks the
symmetry between muon neutrinos and tau neutrinos and creates a
CP-violation dependence for the electron (anti-) neutrino survival
probability and consequently on the electron (anti-) neutrino
flux. Such a phenomenon was numerically observed in
\cite{Balantekin:2007es}. The addition of the non-linear
neutrino-neutrino interaction induces larger effects on the
electron neutrino fluxes up to a level of $10\%$ inside the
supernova \cite{Gava:2008rp}. In this section, we analytically
demonstrate that the inclusion of $V_{\mu \tau}$ (or more
generally the inclusion of an interaction term in the total
Hamiltonian breaking the symmetry between mu and tau neutrinos)
will make the electron survival probability dependent upon the
CP-violation phase $\delta$. The fact that in the SUSY framework,
$V_{\mu \tau}$ can be up to a factor $2 \times \,10^{-2} \,V_e$
implies sizeable effects on the electron neutrino fluxes seen in a
detector on Earth. \setcounter{subsubsection}{0}
\subsubsection{The factorization}
In a dense environment the neutrino evolution equations with
one-loop correction to matter interactions are given by:
\begin{equation}
\label{mswVmutau}
i \frac{\partial}{\partial t} \left(\begin{array}{c} \Psi_e \\ \Psi_{\mu} \\
    \Psi_{\tau}\end{array} \right)  =
     \left[ U \left(\begin{array}{ccc}
     E_1 & 0 & 0  \\
     0 &  E_2  & 0 \\
     0 & 0 &  E_3 \end{array} \right)
U^{\dagger} + \left(\begin{array}{ccc}
     V_c & 0 & 0  \\
     0 &  0  & 0 \\
     0 & 0 &  V_{\mu \tau} \end{array}\right)\right] \left(\begin{array}{c} \Psi_e\\ \Psi_{\mu}
    \\     \Psi_{\tau}\end{array} \right)
\end{equation}

where $ \Psi_{\alpha}$ denotes a neutrino in a flavour state
$\alpha =$e, $\mu,\tau$,  $E_{i=1,2,3}$ being the energies of the
neutrino mass eigenstates, and $U$ the unitary
Maki-Nakagawa-Sakata-Pontecorvo matrix
\begin{equation}
\label{e:2} U = T_{23} T_{13} T_{12} =
 \left(\begin{array}{ccc}
1         &    0       & 0        \\
0         &  c_{23}    & s_{23}     \\
0         & -s_{23}    & c_{23}
\end{array}\right)
 \left(\begin{array}{ccc}
c_{13}                  &    0       & s_{13}\,e^{-i\delta} \\
0                       &    1       & 0       \\
-s_{13}\,e^{i\delta}    &    0       & c_{13}
\end{array} \right)
 \left(\begin{array}{ccc}
c_{12}    &  s_{12}  & 0     \\
-s_{12}   &  c_{12}    & 0     \\
0         &    0       & 1
\end{array} \right) ,
\end{equation}
where $c_{ij} = cos \theta_{ij}$ ($s_{ij} = sin \theta_{ij}$) with
$\theta_{12},\theta_{23}$ and $\theta_{13}$ the three neutrino
mixing angles. The presence of a Dirac $\delta $ phase in
Eq.(\ref{e:2}) renders $U$ complex and introduces a difference
between matter and anti-matter.
 The neutrino interaction with matter is taken into account through
 an effective Hamiltonian which corresponds, at tree level, to the
 diagonal matrix $H_m=diag(V_c,0 ,0)$, where the
 $V_c (x) = \sqrt{2} G_F  N_e (x)$ potential, due to the
 charged-current interaction, depends upon the electron density
 $N_e (x)$ (note that the neutral current interaction introduces
 an overall phase only).
Following the derivation of \cite{Balantekin:2007es,Gava:2008rp},
to obtain explicit relations between probabilities and the
CP-violating phase, it is convenient to work within a new basis
where a rotation by $T_{23}$ is performed. In this basis, one can
factorize the $S$ matrix, defined by $diag(1,1,e^{i\delta})$, out
of the Hamiltonian, so that:
\begin{eqnarray}
\label{mswT23Vmutau}
i \frac{\partial}{\partial t} \left(\begin{array}{c} \Psi_e \\ \tilde{\Psi}_{\mu} \\
    \tilde{\Psi}_{\tau}\end{array} \right)  &=&
    \tilde{H}_T(\delta) \left(\begin{array}{c} \Psi_e\\ \tilde{\Psi}_{\mu}
        \\     \tilde{\Psi}_{\tau}\end{array} \right)
        =S\,\tilde{H'}_T(\delta)\,S^{\dagger}
        \left(\begin{array}{c} \Psi_e\\ \tilde{\Psi}_{\mu}
        \\  \tilde{\Psi}_{\tau}\end{array} \right) \nonumber \\
    &=&
    S \left[ T^0_{13}T_{12} \left(\begin{array}{ccc}
     E_1 & 0 & 0  \\
     0 &  E_2  & 0 \\
     0 & 0 &  E_3 \end{array} \right)
T_{12}^{\dagger}{T^0_{13}}^\dagger
\right. \\
&+&\left. \left(\begin{array}{ccc}
     V_c  & 0 & 0  \\
     0 &  s_{23}^2\,V_{\mu \tau}  & -c_{23}s_{23}e^{i\delta}\,V_{\mu \tau} \\
     0 &  -c_{23}s_{23}e^{-i\delta}\,V_{\mu \tau} &  c_{23}^2\,V_{\mu \tau} \end{array}\right)\right]  S^\dagger \left(\begin{array}{c} \Psi_e\\ \tilde{\Psi}_{\mu}
    \\     \tilde{\Psi}_{\tau}\end{array} \right), \nonumber
\end{eqnarray}
Contrary to the case where only tree level matter interaction is
considered, the $S$ matrix does not commute with the matter
Hamiltonian in the $T_{23}$ basis. This fact implies that: \be
\label{hamredb} \tilde{H}_T(\delta) \neq
S\,\tilde{H}_T(\delta=0)\,S^{\dagger}, \ee and, therefore, that
\be \label{e:13d} \tilde{U}_m(\delta) \neq S \tilde{U}_m(\delta=0)
S^{\dagger}. \ee
\subsubsection{Consequence on the electron neutrino survival probability}
Nevertheless, the factorization of the $S$ matrices is always
possible but the Hamiltonian $\tilde{H'}_T(\delta)$ will depend on
$\delta$. We can rewrite Eq.(\ref{mswT23Vmutau}) in the evolution
operator formalism to be: \be \label{mswT23Vmutaud}
i\frac{\partial}{\partial t} \left(\begin{array}{ccc}
     A_{ee} & A_{\tilde{\mu}e} & A_{\tilde{\tau}e}e^{i\delta}  \\
     A_{e \tilde{\mu}} &  A_{\tilde{\mu}\tilde{\mu}}  & A_{\tilde{\tau}\tilde{\mu}}e^{i\delta} \\
    A_{e \tilde{\tau}}e^{-i\delta} & A_{\tilde{\mu}\tilde{\tau}}e^{-i\delta} &  A_{\tilde{\tau}\tilde{\tau}} \end{array} \right) =\tilde{H'}_T(\delta)\left(\begin{array}{ccc}
     A_{ee} & A_{\tilde{\mu}e} & A_{\tilde{\tau}e}e^{i\delta}  \\
     A_{e \tilde{\mu}} &  A_{\tilde{\mu}\tilde{\mu}}  & A_{\tilde{\tau}\tilde{\mu}}e^{i\delta} \\
    A_{e \tilde{\tau}}e^{-i\delta} & A_{\tilde{\mu}\tilde{\tau}}e^{-i\delta} &  A_{\tilde{\tau}\tilde{\tau}} \end{array} \right)
\ee Defining the Hamiltonian $\tilde{H'}_T(\delta)$
by\footnote{Note that the terms a, b, c, d, e ,f and g are real.}:
\be \tilde{H'}_T(\delta)=\left(\begin{array}{ccc}
     a & b & c  \\
     b &  d  & (e-g\,e^{i\delta}) \\
     c & (e-g\,e^{-i\delta}) &  f \end{array} \right)
\ee we can rewrite the evolution equations for the amplitudes of
the first column of the evolution operator, which corresponds to
the creation of an electron neutrino $\nu_e$ initially:
\begin{eqnarray}
\label{evopAd}
i\frac{d }{dt}A_{ee}&=& a\,A_{ee}+ b\,A_{e \tilde{\mu}} +c\,A_{e \tilde{\tau}}e^{-i\delta} \nonumber \\
i\frac{d }{dt}A_{e \tilde{\mu}}&=& b\,A_{ee}+ d\,A_{e \tilde{\mu}} +(e-g\,e^{i\delta})\,A_{e \tilde{\tau}}e^{-i\delta} \nonumber \\
i\frac{d }{dt}A_{e \tilde{\tau}}e^{-i\delta}&=& c\,A_{ee}+
(e-g\,e^{i\delta})\,A_{e \tilde{\mu}} +f\,A_{e
\tilde{\tau}}e^{-i\delta}
\end{eqnarray}
Similarly, we can write the same equation for the amplitudes
$B_{\alpha\beta}$ when $\delta$ is taken to be zero and look at
the difference between the amplitudes that depend on $\delta$ and
those which do not. The basic idea here is to prove that, because
of the one-loop correction term $V_{\mu\tau}$, the function
($A_{ee}-B_{ee}$) can not remain zero function by showing that its
derivative is non zero.
\begin{eqnarray}
\label{diffevop}
i\frac{d }{dt}(A_{ee}-B_{ee})&=& a\,(A_{ee}-B_{ee})+ b\,(A_{e \tilde{\mu}}-B_{e \tilde{\mu}}) +c\,(A_{e \tilde{\tau}}e^{-i\delta}-B_{e \tilde{\tau}})  \\
i\frac{d }{dt}(A_{e \tilde{\mu}}-B_{e \tilde{\mu}})&=& b\,(A_{ee}-B_{ee})+ d\,(A_{e \tilde{\mu}}-B_{e \tilde{\mu}}) +e\,(A_{e \tilde{\tau}}e^{-i\delta}-B_{e \tilde{\tau}})+g\,(A_{e \tilde{\tau}}-B_{e \tilde{\tau}}) \nonumber \\
i\frac{d }{dt}(A_{e \tilde{\tau}}e^{-i\delta}-B_{e
\tilde{\tau}})&=& c\,(A_{ee}-B_{ee})+ e\,(A_{e \tilde{\mu}}-A_{e
\tilde{\mu}}) +f\,(A_{e \tilde{\tau}}e^{-i\delta}-B_{e
\tilde{\tau}})-g\,(A_{e \tilde{\mu}}e^{-i\delta}-B_{e
\tilde{\mu}}) \nonumber
\end{eqnarray}
Let us now take a closer look at Eqs.(\ref{diffevop}). The initial
condition we are interested in, namely a $\nu_e$ created initially
means that initially the amplitudes $A$ and $B$ are: \be
\left(\begin{array}{c} \Psi_e\\ \Psi_{\mu}
        \\  \Psi_{\tau}\end{array} \right)= \left(\begin{array}{c} 1 \\ 0 \\ 0 \end{array} \right)
\Rightarrow \left(\begin{array}{c} A_{ee}\\ A_{e \tilde{\mu}}
        \\  A_{e \tilde{\tau}}e^{-i\delta}\end{array} \right)=
        \left(\begin{array}{c} B_{ee}\\ B_{e \tilde{\mu}}
        \\  B_{e \tilde{\tau}}\end{array} \right)=
        \left(\begin{array}{c} 1 \\ 0 \\ 0 \end{array} \right)
 \ee
When $g=0$ ($V_{\mu \tau}=0$), it is easy to see that inserting
the initial conditions into Eqs.(\ref{diffevop}) will imply that
the functions  $f_e=A_{ee}-B_{ee}$, $f_{\mu}=A_{e
\tilde{\mu}}-B_{e \tilde{\mu}}$ and $f_{\tau}=A_{e
\tilde{\tau}}e^{-i\delta}-B_{e \tilde{\tau}}$ will be equal to
zero. By discretizing time we see, by recurrence, that if those
functions are zero at beginning, they will be equal to zero at all
time. But when $g \neq 0$ ($V_{\mu \tau}\neq0$) we have to look
also at the evolution of the functions $\hat{f}_{\mu}=A_{e
\tilde{\mu}}e^{-i\delta}-B_{e \tilde{\mu}}$ and
$\hat{f}_{\tau}=A_{e \tilde{\tau}}-B_{e \tilde{\tau}}$. Their
respective evolution equation can be easily derived from
Eq.(\ref{evopAd}) to yield: \be i\frac{d }{dt}(A_{e
\tilde{\mu}}e^{-i\delta}-B_{e \tilde{\mu}})=
b\,(A_{ee}e^{-i\delta}-B_{ee})+ d\,(A_{e
\tilde{\mu}}e^{-i\delta}-B_{e \tilde{\mu}}) +e\,(A_{e
\tilde{\tau}}e^{-2i\delta}-B_{e \tilde{\tau}})+g\,(A_{e
\tilde{\tau}}e^{-i\delta}-B_{e \tilde{\tau}}) \ee \be i\frac{d
}{dt}(A_{e \tilde{\tau}}-B_{e \tilde{\tau}})=
c\,(A_{ee}e^{i\delta}-B_{ee})+ e\,(A_{e
\tilde{\mu}}e^{i\delta}-A_{e \tilde{\mu}}) +f\,(A_{e
\tilde{\tau}}-B_{e \tilde{\tau}})-g\,(A_{e \tilde{\mu}}-B_{e
\tilde{\mu}}) \ee Initially, the derivatives are :
\begin{eqnarray}
i\frac{d }{dt}(A_{e \tilde{\mu}}e^{-i\delta}-B_{e \tilde{\mu}})(t=0)&=& i\frac{d}{dt}\hat{f}_{\mu}(t=0)= b(e^{-i\delta}-1) \nonumber \\
i\frac{d }{dt}(A_{e \tilde{\tau}}-B_{e \tilde{\tau}})(t=0)&=&
i\frac{d}{dt}\hat{f}_{\tau}(t=0)=c(e^{i\delta}-1)
\end{eqnarray}
We just proved that since the functions $\hat{f}_{\mu}$ and
$\hat{f}_{\tau}$ are non constant zero functions, the functions
$f_{\mu}$ and $f_{\tau}$ won't be zero as well. But does it
implies that the function $f_e$ is non zero at all time? No,
because the contributions from $\hat{f}_{\mu}$ and
$\hat{f}_{\tau}$ could cancel in the evolution equation
(\ref{diffevop}) of $f_e$. To precisely study the evolution of
$f_e$, we discretize time such as $t=N*\Delta t$ with $N \in
\mathbb{N}$ and \be \frac{d}{dt}f_e =\frac{f_e(t+\Delta
t)-f_e(t)}{\Delta t} \ee Using Eqs.(\ref{diffevop}) and the time
discretization we see that: At $t=\Delta t$:
\begin{eqnarray}
\hat{f}_\mu(\Delta t)&=& \frac{b(e^{-i\delta}-1)}{i}\Delta t\nonumber \\
\hat{f}_{\tau}(\Delta t)&=&\frac{c(e^{i\delta}-1)}{i}\Delta t \nonumber \\
\end{eqnarray}
which implies that: At $t=2\Delta t$:
\begin{eqnarray}
f_\mu(2\Delta t)&=& gb(e^{-i\delta}-1)\Delta^2 t \nonumber \\
f_{\tau}(2\Delta t)&=& -gc(e^{i\delta}-1)\Delta^2 t \nonumber \\
\end{eqnarray}
leading at $t=3\Delta t$
\begin{eqnarray}
f_e(3\Delta t)&=&\frac{1}{i}\left(gbc(e^{-i\delta}-1)-gbc(e^{i\delta}-1) \right)\Delta^3 t \nonumber \\
&=& \frac{1}{i}gbc(e^{-i\delta}-e^{i\delta})\Delta^3 t \nonumber \\
&=& -2gbc \sin \delta \Delta^3 t
\end{eqnarray}
This last formula proves that the function $f_e$ is not the
constant zero function when $\delta \neq 0$\footnote{Note that
with such formula, $f_e$ is also equal to zero when $\delta= \pi$,
but going to the forth step will show, after a tedious but
straightforward calculation, that $f_e$ is non zero even when
$\delta=\pi$.}, therefore $A_{ee} \neq B_{ee}$ and consequently,
for $V_{\mu\tau} \neq 0$: \be \label{PnuenueVmutau} P (\nu_{e}
\rightarrow \nu_e, \delta \neq 0) \neq P (\nu_{e} \rightarrow
\nu_e, \delta = 0) \ee Therefore, when $\delta$ is non zero, it
has an influence on the value of $P (\nu_{e} \rightarrow \nu_e,
\delta)$. The luminosity of a neutrino emitted initially as a
flavour $\alpha$ is
 \begin{equation}\label{e:4quadris}
L_{{\nu}_{\underline{\alpha}}}(r,E_{\nu})=
{L^0_{{\nu}_{\underline{\alpha}}}
\over{T_{\nu_{\underline{\alpha}}}^3 \langle
E_{\nu_{\underline{\alpha}}} \rangle
F_2(\eta)}}{{E_{\nu_{\underline{\alpha}}}^2} \over{1 +
\exp{(E_{\nu_{\underline{\alpha}}}/T_{\nu_{\underline{\alpha}}} -
\eta)}}}
\end{equation}
where $F_2(\eta)$ is the Fermi integral,
$L^0_{{\nu}_{\underline{\alpha}}}$ and
$T_{\nu_{\underline{\alpha}}}$ are the luminosity and temperature
at the neutrinosphere. The $\nu_e$ and $\bar{\nu}_e$ fluxes will
depend on $\delta$ even when the luminosities
$L_{\nu_{\underline{\mu}}}$ and $L_{\nu_{\underline{\tau}}}$ are
taken equal at the neutrino-sphere:
\begin{eqnarray}
\label{fluxdelta2} {\phi}_{\nu_e}(\delta) &=&
L_{\nu_{\underline{e}}}P(\nu_e \rightarrow \nu_e,\,\delta) +
L_{\nu_{\underline{\mu}}} \left( P(\nu_\mu \rightarrow \nu_e)+P(\nu_\tau \rightarrow \nu_e) \right) \nonumber \\
&=&L_{\nu_{\underline{e}}}P(\nu_e \rightarrow \nu_e,\,\delta) +
L_{\nu_{\underline{\mu}}} \left( 1- P(\nu_e \rightarrow \nu_e,\,\delta) \right)  \nonumber \\
&=& (L_{\nu_{\underline{e}}}-L_{\nu_{\underline{\mu}}})P(\nu_e
\rightarrow \nu_e,\,\delta) + L_{\nu_{\underline{\mu}}}
\end{eqnarray}
This analytical derivation proves that, if the dependence on
$\delta$ of the evolution operator cannot be factorized then the
electron neutrino survival probability depends on $\delta$.
Nevertheless, this implication had been observed numerically. We
can easily generalize this derivation to other interactions that
would distinguish $\nu_\mu$ from $\nu_\tau$ like for instance,
non-standard neutrino interaction \cite{EstebanPretel:2007yu}.
Therefore we can state that as soon as the medium effect on
$\nu_\mu$ and on $\nu_\tau$ is not the same, effects of the
CP-violating phase on the electron neutrino (and anti-neutrino)
fluxes will appear.

\section{Conclusions}
In this paper, we have investigated the radiative correction on
the $\mu -\tau$ neutrino indices of refraction coming from beyond
standard physics. In the NMSSM, we have shown that the sign of
$V_{\mu \tau}$ depends upon the hierarchy of the sleptons masses
and therefore could be negative contrary to the Standard Model
case. After writing and adding a subroutine to a low-energy code
taking into account all current constraints on SUSY we showed that
$\varepsilon$ can increase up to the order of $2\times10^{-2}$
depending on the supersymmetric parameters. Such value could be
highly important in the calculation of neutrino fluxes from
core-collapse supernovae as it can induce sizeable effects upon
the electron (anti)-neutrino fluxes. In a second part we have
calculated all contributions from R-parity breaking interactions
on the radiative corrections $V_{\mu \tau}$. Taking into account
such interactions, we showed that NMSSM distinguishes from MSSM in
this case and bring new possible contributions. The next step in
this type of calculations would be to see the consequences of
theses values of $V_{\mu \tau}$ on the supernova neutrino fluxes
and to use it in order to survey the supersymmetric parameter
space \cite{CCJerome}. Secondly, we would have to calculate
corrections with gravitino loops and all contributions from
R-parity breaking interactions. Another interesting possibility
would be to calculate the radiative corrections for the
neutrino-neutrino interaction. Such calculation has been recently
done in the SM \cite{Mirizzi:2009td} and SUSY framework could
yield potentially much bigger effect. Finally, as an application
for $V_{\mu \tau}$ we demonstrated that the inclusion of such a
term implies that the electron (anti-) neutrino survival
probability, and consequently the electron (anti-) neutrino
fluxes, will depend upon $\delta$. The consequences of such
dependence and the fact that $V_{\mu \tau}$ can be up to
$2\times10^{-2} \,V_e$ will be studied in a future work.

\section*{Acknowledgements}

The authors are grateful to S. Friot for his enthusiasm and
encouragement during the realization of this work and for useful
discussion. They also want to thank  F. Domingo, J. Kneller and S.
Descotes-Genon for fruitful discussions. Finally, they thank A.
Mirizzi, P. D. Serpico and J. Kneller for the careful reading of
the manuscript.

\textit{} 

\end{document}